\pdfoutput=1
\documentclass[11pt]{article}
\usepackage{jheppub}
\usepackage{graphicx,amsmath,amssymb,amsthm}
\usepackage{hyperref}
\usepackage[dvipsnames]{xcolor}
\usepackage{thmtools, thm-restate}
\usepackage{epsfig}
\usepackage{epstopdf}
\usepackage{latexsym}
\usepackage{graphicx}
\usepackage{booktabs}
\usepackage{bbm}
\usepackage{color}
\usepackage{physics}
\usepackage{tensor}
\usepackage{verbatim}
\usepackage[caption=false]{subfig}
\usepackage{tikz}
\usepackage{bigints}
\usepackage{ifthen}
\usetikzlibrary{matrix}
\usetikzlibrary{decorations.markings,calc,shapes,decorations.pathmorphing}
\usetikzlibrary{patterns}
\usetikzlibrary{positioning}
\usepackage{textpos}

\hypersetup{
colorlinks,
linkcolor={blue!80!black},
citecolor={blue!80!black},
urlcolor={blue!80!black},
linktoc=page
}

\begin{document}
\def\nn{\nonumber}
\def\kc#1{\left(#1\right)}
\def\kd#1{\left[#1\right]}
\def\ke#1{\left\{#1\right\}}
\newcommand\beq{\begin{equation}}
\newcommand\eeq{\end{equation}}
\renewcommand{\Re}{\mathop{\mathrm{Re}}}
\renewcommand{\Im}{\mathop{\mathrm{Im}}}
\renewcommand{\b}[1]{\mathbf{#1}}
\renewcommand{\c}[1]{\mathcal{#1}}
\renewcommand{\u}{\uparrow}
\renewcommand{\d}{\downarrow}
\newcommand{\be}{\begin{equation}}
\newcommand{\ee}{\end{equation}}
\newcommand{\bsigma}{\boldsymbol{\sigma}}
\newcommand{\blambda}{\boldsymbol{\lambda}}
\newcommand{\sgn}{\mathop{\mathrm{sgn}}}
\newcommand{\diag}{\mathop{\mathrm{diag}}}
\newcommand{\Pf}{\mathop{\mathrm{Pf}}}
\newcommand{\half}{{\textstyle\frac{1}{2}}}
\newcommand{\sh}{{\textstyle{\frac{1}{2}}}}
\newcommand{\ish}{{\textstyle{\frac{i}{2}}}}
\newcommand{\thf}{{\textstyle{\frac{3}{2}}}}
\newcommand{\SUN}{SU(\mathcal{N})}
\newcommand{\N}{\mathcal{N}}

\newcommand{\psp}{r_{\text{ps}}}

\newcommand{\haoyu}{\textcolor{blue}}
\newcommand{\mk}{\textcolor{red}}

\title{The Photon Sphere and the AdS/CFT Correspondence}

\abstract
{The AdS/CFT correspondence connects bulk fields $\phi$ to boundary operators $\mathcal{O}$ characterized by source frequency $\omega$ and angular momentum $l$. Here we explore their connection to massless particles with an impact parameter $b=\omega/l$. In the AdS Schwarzschild spacetime, these particles follow unstable orbits around the photon sphere -- with Lyapunov exponent $\lambda$ -- when $b$ is near a critical value. The behavior of the bulk field is obtained numerically and then studied using an analytic approach, which leads to a precise approximate formula for the amplitude of the bulk field $\phi$. This gives the correct qualitative behavior for the system, with the amplitude of the field taking the shape of an arrowhead in tortoise coordinates. The field behaves analogously to the massless particles, and the amplitude of $\phi$ diverges at the critical impact parameter when the source frequency takes the value $\omega \approx \lambda l$, where $\lambda$ is the Lyapunov exponent of the null geodesics. We find this transition occurs when $b = \lambda$. We show this is precisely when the first QNM becomes available, and obtain an approximate formula for the first few overtones.  }

\author{Marcos Riojas$^{a}$, Hao-Yu Sun$^{a}$}
\affiliation{$^a$Weinberg Institute, Department of Physics, the University of Texas at Austin, TX 78712-1192, USA.}
\emailAdd{marcos.riojas@utexas.edu}
\emailAdd{hkdavidsun@utexas.edu}

\maketitle

\section{Introduction}

For many first-time learners of general relativity, the photon sphere, originally examined in \cite{Bardeen:1972fi} as marginally trapped light rays, may appear as a seemingly inconsequential peculiarity associated with black holes, as commonly presented in textbooks such as \cite{Misner:1973prb,hartle2003gravity}. However, it is important to note that photon spheres actually play a significant role in various aspects of astrophysics and gravitational physics, demonstrating their profound importance in understanding the properties and behavior of black holes. They have recently gained significant attention for two compelling reasons. Firstly, on the more theoretical side, they play a crucial role in imaging Einstein rings of black holes in Anti-de Sitter (AdS) spacetime \cite{Hashimoto:2018okj,Hashimoto:2019jmw,Liu:2022cev}, in the emergence of conformal symmetries associated with enhanced bright regions (``photon (sub)rings'') from gravitaional lensing by a Kerr black hole\footnote{To be more precise, there is no ``photon sphere'' in this case, but rather a \emph{photon region} \cite{Perlick:2004tq,Grenzebach:2014fha,Grenzebach:2015oea}, or sometimes called 
a \emph{photon shell} \cite{Johnson:2019ljv}.} in asymptotically flat spacetime \cite{Johnson:2019ljv,Gralla:2019drh,Himwich:2020msm,Hadar:2022xag}, and in other theoretical implications such as warped AdS$_3$ black holes \cite{Kapec:2022dvc}. Secondly, the practical application of imaging astrophysical black hole shadows, such as those of M87* \cite{EventHorizonTelescope:2019dse} and Sgr A* \cite{EventHorizonTelescope:2022wkp} by the Event Horizon Telescope, has sparked renewed interest in the study of photon spheres. An extensive review on analytical techniques for calculating black hole photon rings and shadows can be found in \cite{PERLICK20221}.

In addition to photon spheres, another key player in the study of black holes is the concept of quasi-normal modes (QNMs). For scalar fields, QNMs are roughly defined as solutions to the Klein--Gordon equation \eqref{eq:KG} that are purely outgoing near infinity ($\Phi\sim e^{-i\omega(t-r_*)}$) and purely ingoing near the black hole horizon ($\Phi\sim e^{-i\omega(t+r_*)}$). Only a discrete set of complex frequencies $\omega$ is allowed due to the absence of physical meaning of the initial incoming waves from infinity. In asymptotically AdS spacetimes, the effective potential (see Appendix \ref{sec:WKB}) vanishes exponentially near the horizon and diverges at infinity, requiring $\Phi(r)$ to vanish at infinity. Complex frequencies determine the late-time fall-off of fields $\Phi(r)$.

Albeit just being solutions to linear second-order ODEs, QNMs offer a powerful approach for understanding the behavior of perturbations around black holes \cite{Berti:2009kk}. Various techniques have been developed to solve for QNMs over the past few decades. The WKB method, along with its variant called the monodromy method for highly damped modes, has proven to be effective in many types of spacetime, particularly asymptotically flat ones \cite{Schutz:1985km,Iyer:1986np,Iyer:1986nq}. Frobenius series solutions are well-suited for asymptotically AdS spacetimes, while a resonance method has shown better results for small black holes. The use of continued fractions, first introduced by Leaver \cite{Leaver:1985ax,leaver1986solutions,Leaver:1986gd}, has also been applied to asymptotically AdS spacetimes recently \cite{Daghigh:2022uws}. Other methods include Regge poles \cite{Decanini:2009dn} and purely numerical techniques. Numerous investigations of QNMs have been subsequently conducted, as summarized in the comprehensive review by Berti et al. \cite{Berti:2009kk}. Several important aspects contribute to the understanding of QNMs. The retarded Green's function and the fact that they do not form a complete set of basis \cite{Berti:2009kk} are fundamental in their calculations. The computation involves an inverse Laplace transform, whose contour integral on the complex frequency plane includes contributions from two quarter-circles, a branch cut responsible for late-time long tails\footnote{For both asymptotically flat and AdS spacetimes, see e.g., \cite{Price:1972pw,Ching:1994bd,Ching:1995tj}.}, and a sum over residues. For recent computational results using 4d $\mathcal{N}=2$ supersymmetric gauge theories, see e.g., \cite{Aminov:2020yma} and \cite{Bonelli:2021uvf}.

Early studies of black holes in asymptotically AdS spacetime were conducted by Horowitz and Hubeny in \cite{Horowitz:1999jd} among others, revealing important properties such as that QNM frequencies (both their real and complex parts) are proportional to the temperatures of large black holes. Notably, contrary to QNMs for the black holes in asymptotically flat spacetimes, where \cite{Price:1972pw} revealed the existence of power-law tails, those of AdS-Schwarzschild black holes do not exhibit such tails as shown by \cite{Ching:1994bd,Ching:1995tj}. Power-law behavior in the intermediate-time regime of QNMs is also absent in AdS-Schwarzschild \cite{Horowitz:1999jd}.

Moreover, QNMs have further deepened the understanding of the AdS/CMT (condensed matter theory) correspondence \cite{hartnoll2018holographic}, because they are natural physical quantities in the dual strongly coupled field theory without weakly interacting quasiparticle excitations. Specifically, when going beyond the scalar operators $\mathcal{O}$ to current operators $J^\mu$, QNMs also describe the charge transport in quantum critical theory. Moreover, a functional determinant or partition function for semiclassical gravity in both asymptotically AdS and dS can be expressed as a product over all QNMs \cite{Denef:2009kn,Keeler:2014hba,Keeler:2016wko}, offering $1/N$ corrections, such as the singular long-time tail in electrical conductivity at the one-loop level \cite{Kovtun:2003vj}. Other related applications of QNMs include determining the viscosity-to-entropy-density ratio in hyperscaling-violating Lifshitz theories \cite{Mukherjee:2017ynv}, as well as studying strange metals \cite{Andrade:2018gqk}, hydrodynamic transports \cite{Policastro:2002tn,Herzog:2003ke,Kovtun:2005ev}, operator mixings \cite{Kaminski:2009dh}, and phase transitions in superfluids and superconductors \cite{Amado:2009ts,Bhaseen:2012gg}.

It should be noted that the relationship between QNMs and the size of black hole photon spheres (or shadows) is not simply expressed in terms of the impact parameter in AdS spacetimes\footnote{They are in both asymptotically flat and dS spacetimes.}, as discussed in \cite{PERLICK20221} and \cite{Cardoso:2008bp}. This intricacy adds to the subtlety of the physical connections between QNMs and photon spheres in AdS spacetime. Moreover, recent research by Bardeen \cite{Bardeen:2018omt} has highlighted the predominant generation of Hawking radiation near the photon sphere of a black hole in asymptotically flat spacetime. This finding has reemphasized the importance of understanding the relationship between QNMs and photon spheres. Further motivated by the intriguing way of detecting the existence of a gravitational dual of a given $3$d boundary theory using the temperature-dependence of $4$d planar AdS-Schwarzschild black hole images \cite{Hashimoto:2018okj,Hashimoto:2019jmw}, we delve deeper into the precise relationship between photon spheres and QNMs here. 

We exclusively work with massless fields $\Phi(t,r,\theta,\phi)$ in the AdS$_4$-Schwarzschild background, satisfying the Klein--Gordon equation. In the light of AdS/CFT correspondence, its conformal dimension dictates its asymptotic behavior near the AdS boundary -- its corresponding oscillating source ($\propto e^{i\omega t}$) and the expectation value of its dual operator can be read off from the leading and subleading terms, respectively. When solving the Klein--Gordon equation, we obtain a Helmholtz-like equation for radial modes $\phi_l(r)$ labeled by angular momenta, saying that for large $l$, $\phi_l(r)$ are pushed away from the origin, where the effective potential takes its maximum value. 

The effective potential takes its local maximum values at two places: at the AdS boundary, and at the photon sphere. The solutions to the Klein--Gordon equation are well approximated by null geodesics in the WKB limit, so we expect the field to be confined between the AdS boundary and the photon sphere for a high enough angular momentum $l$. Since the scalar field is related to the source and the expectation value of the operator it sources, we suggest the AdS/CFT correspondence could depend in an interesting way on the angular momentum of the dual operator.

By exploring an interesting connection between the Klein--Gordon equation and the Schrodinger equation, we then derive a formula that describes the amplification of fields as they propagate from the UV cut-off to the horizon of the black hole \eqref{eq:amplification_formula}. Using the WKB approximation and a field redefinition, the formula has a physical interpretation as a tunneling process. The derivation then follows from the cutoff invariance of the theory. For black holes of the order of the Hawking--Page transition and above, the formula gives an excellent approximation for the amplitude of a field at the horizon of the black hole when the eikonal approximation is valid. We then explain the breakdown of this approximation in terms of massless particles orbiting the photon sphere many times in the vicinity of their critical impact parameter. Finally, we use similar reasoning to determine the QNMs of small (below the Hawking--Page temperature) AdS black holes in the large angular momentum limit.

The rest of the paper is organized as follows. In Section \ref{sec:photon}, we review the notion of a photon sphere in AdS spacetime in the context of classical particles, set up notation and coordinate systems, and describe the behavior of massless particles traveling along null geodesics in terms of an impact parameter. In Section \ref{sec:KG} we set up an analogous problem for the Klein--Gordon equation. In Section \ref{sec:amplification}, we derive the amplification formula \eqref{eq:amplification_formula}, which determines the fate -- amplifying or decaying -- of a signal incident from the asymptotic boundary as it approaches an AdS black hole. It also predicts two phase transitions in accordance with the geodesic approximation in the eikonal limit. In Section \ref{sec:Lyapunov}, we compute the Lyapunov exponent, which turns out to equal the critical value of the impact parameter. We then determine the QNMs of an AdS black hole, well below the Hawking--Page transition, at high angular momentum $l$. We obtain a closed-form analytical solution that depends on the Lyapunov exponent, but we obtain the modes using an approximation method outlined in \cite{Cardoso:2008bp}. In Section \ref{sec:numerics}, we present our numerical results in the context of the amplification formula and speculate on the meaning of our findings. We conclude in Section \ref{sec:conclusion}, leaving the full investigation on the large-angular-momentum sector in the dual CFT inspired by the photon sphere to later, among other future directions. We collected various technical details on solving ODEs and a review on the WKB approximation to Appendices \ref{sec:AppendixA} and \ref{sec:WKB}. Finally, Appendix \ref{sec:figures} contains many visualizations of the amplification formula \eqref{eq:amplification_formula}.

\section{The Photon Sphere}
\label{sec:photon}

This section is an overview of known properties of the photon sphere, generalizing them to the AdS case when necessary. The photon sphere plays a well-known role in the determination of quasinormal modes in asymptotically flat spacetimes \cite{Cardoso:2008bp, Decanini:2009dn},  and is an important astrophysical object in gravitational wave astronomy \cite{PERLICK20221, Berti:2009kk}. This discussion closely follows a review of black hole shadows\footnote{The shadow of a black hole is generally \emph{larger} than its photon sphere. This can be roughly seen in Figure \ref{fig:geodesics}. The former is defined by light rays entirely outside of the latter, and \emph{actually} reach a distant observer; see, e.g., \cite{Falcke:1999pj}.} \cite{PERLICK20221}, which are closely related to the photon sphere. Many of these concepts were originally described in \cite{Bardeen:1972fi}. 

\subsection{Conserved Quantitites and the Impact Parameter}

These arguments apply to spherically symmetric and static metrics: 
\begin{equation}
g_{\mu \nu} d x^\mu d x^\nu=-A(r) d t^2+B(r) d r^2+D(r)\left(d \vartheta^2+\sin ^2 \vartheta d \varphi^2\right),
\end{equation}
with $A(r),B(r),D(r)>0$. One key parameter is the amount of angular momentum needed by a massless particle to travel between two points on the AdS boundary. The Lagrangian $\mathcal{L}(x, \dot{x})=(1 / 2) g_{\mu \nu} \dot{x}^\mu \dot{x}^\nu$ permits a particularly useful form that eliminates the square root:
\begin{equation}
\mathcal{L}(x, \dot{x})=\frac{1}{2}\left(-A(r) \dot{t}^2+B(r) \dot{r}^2+D(r)\left(\dot{\vartheta}^2+\sin ^2 \vartheta \dot{\varphi}^2\right)\right).
\end{equation}
This system is axially symmetric, so we choose a convenient angle to work with -- i.e., the equatorial plane: $\vartheta = \frac{\pi}{2}$. The Lagrangian does not depend on $\vartheta$ or $t$, so the equations of motion
\begin{equation}
\frac{d}{d \lambda}\left(\frac{\partial \mathcal{L}}{\partial \dot{x}^\mu}\right)-\frac{\partial \mathcal{L}}{\partial x^\mu}=0,
\end{equation}
yield the energy ($\omega$) and angular momentum ($l$) as conserved quantities: 
\begin{equation}
\omega=A(r) \dot{t}, \quad l=D(r) \dot{\varphi}.
\label{EQ:conserved_quantities}
\end{equation}
The ratio of these conserved quantities is called the ``impact parameter":\footnote{There are differing conventions -- some articles choose $b = \frac{\omega}{l}$, and some choose $b = \frac{l}{\omega}$.} 
\begin{equation}
    b := \frac{\omega}{l}.
    \label{eq:geodesic_impactparameter}
\end{equation}
Its value determines the fate of a particle emitted from the asymptotic boundary. 

\subsection{The Effective Potential and the Impact Parameter}

The first integral of the geodesic equation for massless particles is a useful quantity:
\begin{equation}
-A(r) \dot{t}^2+B(r) \dot{r}^2+D(r) \dot{\varphi}^2=0 .
\label{eq:general_metric}
\end{equation}
Rewriting this using Equation \eqref{EQ:conserved_quantities}, and then solving for $\frac{\dot{r}^2}{\dot{\varphi}^2} = \left(\frac{dr}{d\varphi}\right)^2$ gives:
\begin{equation}
\left(\frac{d r}{d \varphi}\right)^2=\frac{D(r)}{B(r)}\left(\frac{D(r)}{A(r)} \frac{\omega^2}{l^2}-1\right) .
\label{EQ:geodesic_equation}
\end{equation}
The expression in Equation \eqref{EQ:geodesic_equation} depends on the coefficients in the metric, $A(r)$, $B(r)$, and $D(r)$, and on the ratio $b$ \eqref{eq:geodesic_impactparameter} of the conserved quantities $\omega$ and $l$. 

Equation \eqref{EQ:geodesic_equation} takes the form of an oscillator. Suppose some geodesic leaves and returns to the boundary, with a turning point in the trajectory at some radial location $R_t$. At this point we have $dr/d\phi =0$, so $R_t$ and $b$ are related\footnote{Some sources take the reciprocal of this definition.}: 
\begin{equation}
    \frac{\omega^2}{l^2} = b^2 = \frac{A(R_t)}{D(R_t)}.
\end{equation}
It is convenient to introduce a function called the effective potential\footnote{Following a coordinate change $u = 1/r$, this is equivalent to the function $h(u)$ used in previous work by one of the authors \cite{Geng:2020fxl, Geng:2021hlu,Karch:2023ekf}.}: 
\begin{equation}
    h(r) := \frac{A(r)}{D(r)},
    \label{eq:effective_potential}
\end{equation} 
which implies that the turning point is related to the impact parameter as
\begin{equation}
    b^2 = h(R_t).
    \label{eq:impact}
\end{equation}
The equations of motion can then be written as: 
\begin{equation}
\left(\frac{d r}{d \varphi}\right)^2=\frac{D(r)}{B(r)}\left(\frac{h(r)}{h(R_t)}-1\right).
\end{equation}
The discussion up to this point has been mostly general, though spherical symmetry has been assumed. Partly specializing to a Schwarzschild black hole by taking $A(r) = \frac{1}{B(r)}$, we get: 
\begin{equation}
g_{\mu \nu} d x^\mu d x^\nu=-A(r) d t^2+ \frac{1}{A(r)} d r^2+D(r) \left(d \vartheta^2+\sin ^2 \vartheta d \varphi^2\right).
\end{equation}
Using Equation \eqref{EQ:geodesic_equation} and Equation \eqref{eq:impact}, we obtain an expression that can be used to determine the classical turning points for the motion:
\begin{equation}
\begin{aligned}
\left(\frac{d r}{d \varphi}\right)^2=\frac{D(r)}{B(r)}\left(\frac{D(r)}{A(r)} \frac{\omega^2}{l^2}-1\right) & =D(r)^2 \left(\frac{\omega^2}{l^2}-\frac{A(r)}{D(r)}\right) \\
& = D(r)^2 \left( b^2 - h(r) \right).
\end{aligned}
\label{eq:classical_turning_points}
\end{equation}
Massless particles traveling along geodesics at position $R$ travel toward (away) from the boundary depending on how their effective potential compares to their impact parameter. Regardless of where they originate -- whether deep in the bulk, or at the asymptotic boundary -- if they can reach that point, they will turn around at radial position $h(R_t) = b^2$.

\subsection{The Photon Sphere and Geodesic Orbits in AdS-Schwarzschild}

\begin{figure}
    \centering
    \includegraphics[width=\linewidth]{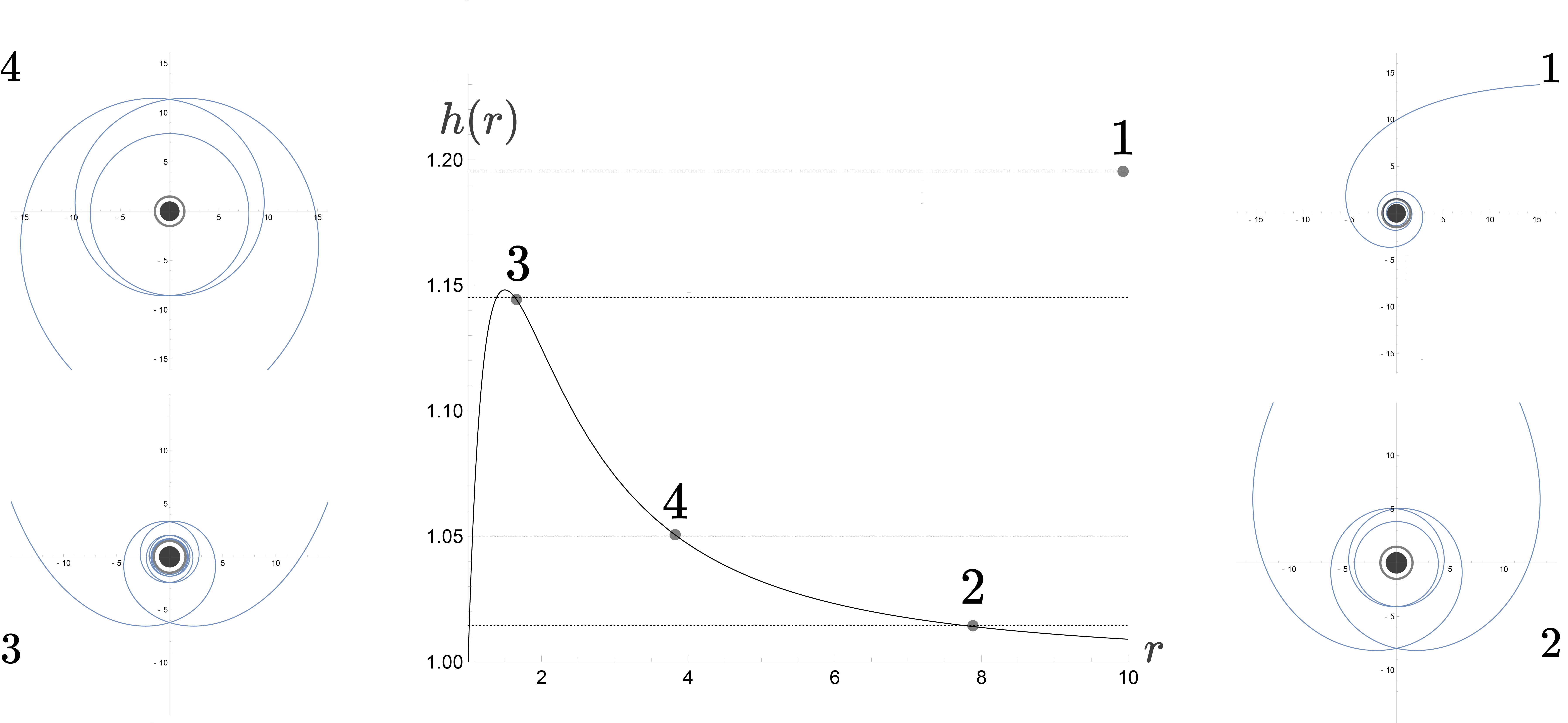}
    \caption{Here we show the geodesic trajectories for massless particles in an AdS spacetime. The photon sphere is pictured as the dark ring around the horizon in black. Their behavior can be predicted from their impact parameters alone from Equation \eqref{eq:geodesiccross}. Panel $1$ shows an inspiral trajectory because the value of its impact parameter is greater than the critical value at the photon sphere, i.e., $b > 1 + 1/27M^2$. The other three panels show scattering trajectories. Particles with impact parameters $b < 1$ cannot leave the asymptotic boundary and are not pictured. One can obtain a scattering (or inspiral) trajectory that orbits the photon sphere arbitrarily many times before returning to the boundary (or falling into the black hole) by choosing an impact parameter sufficiently close to the critical value. We find that the solutions to the scalar wave equation \eqref{eq:Helmholtzform} behave similarly; modes corresponding to sources of frequency $\omega$, which drive operators of angular momentum $l$, penetrate the photon sphere only when their ``impact parameter" satisfies essentially the same condition \eqref{eq:amplification_formula}. Moreover, the solutions are unstable, sharply peak at the photon sphere, and violate \eqref{eq:amplification_formula} when $b$ is sufficiently close to the critical value. See Figures \ref{fig:HP_amplitudes}-\ref{fig:tiny_amplitudes}. }
    \label{fig:geodesics}
\end{figure}

Further specializing to an AdS-Schwarzschild black hole in asymptotically AdS$_d$ spacetime: 
\begin{equation}
g_{\mu \nu} d x^\mu d x^\nu=-F(r) d t^2+ \frac{1}{F(r)} d r^2+  r^2 d\Omega_{d-2},
\end{equation}
and using Equation \eqref{eq:general_metric}, one finds the blackening function, $F(r)$, the effective potential $h(r)$ \eqref{eq:effective_potential}, and the equations of motion \eqref{eq:classical_turning_points} are:
\begin{equation}
\label{eq:effpot}
\begin{aligned}
    F(r) &= 1 + \frac{r^2}{L^2} - \frac{2 M}{r^{d-3}},\\
    h(r) &= \frac{F(r)}{r^2},
    \end{aligned}
\end{equation}
\begin{equation}
    \hspace{-0.2cm}\left(\frac{d r}{d \varphi}\right)^2 = r^2 \left( b^2 - \frac{F(r)}{r^2} \right).
    \label{eq:EOMparticle}
\end{equation}
The behavior of the effective potential is illustrated in Figure \ref{fig:effective_potential_radial}. The orbits of the null geodesics are of particular interest. The boundary conditions are found by initializing $r=r_0$ and determining the momentum from \eqref{eq:EOMparticle}: 
\begin{equation}
    p = \frac{dr}{d\tau} = \left(\frac{dr}{d\varphi}\right)\left( \frac{d\varphi}{d\tau}\right) = \sqrt{\left( \omega^2 - \frac{ l^2 F(r)}{r^2}\right)}.
\end{equation}
One then solves \eqref{eq:EOMparticle} to obtain the null geodesics, which were computed in \textit{Mathematica} for AdS spacetime and illustrated in Figure \ref{fig:geodesics}. There is an unstable point where massless particles can travel around the black hole; their trajectories form a geometric object known as the \textit{photon sphere}. Its radius $r_{ps}$ is an unstable maximum of the effective potential: 
\begin{figure}
    \centering
    \includegraphics[width=.7\linewidth]{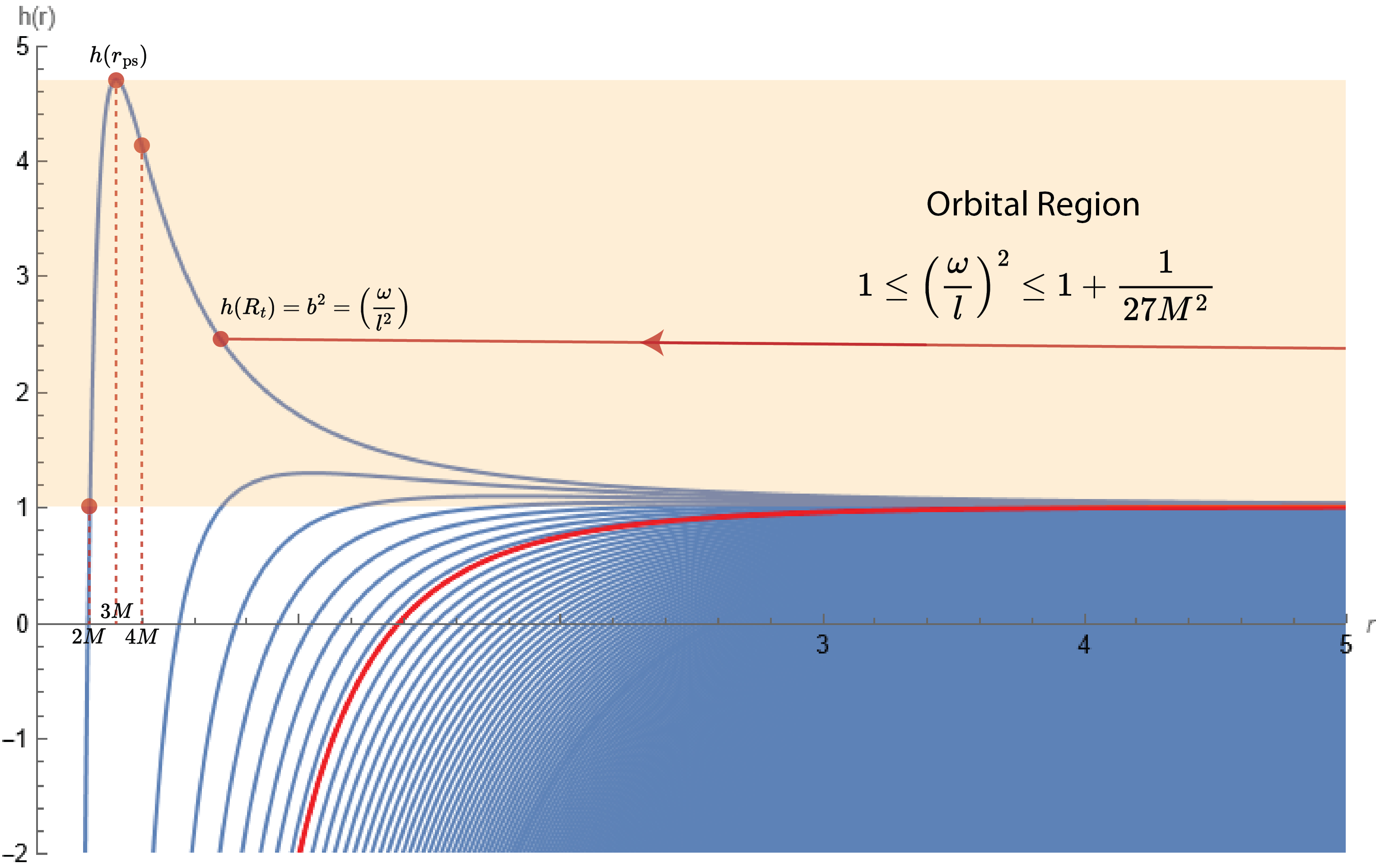}
    \caption{Here we show the effective potential, $h(r) = F(r)/r^2$, for a massless particle near  an asymptotically AdS-Schwarzschild black hole for a large range of black hole masses $M \in (0.1, 200)$. The effective potential decreases monotonically with $M$. For a particle to leave the AdS boundary and travel between boundary points without falling into the black hole, its impact parameter must lie within $h(2M)<b^2<h(3M)$, corresponding to the shaded region for $M=0.1$. The red line corresponds to the Hawking--Page transition that occurs when the radius $R$ of the black hole becomes smaller than the AdS radius $L$. The effective potential has a clear maximum (at the photon sphere) that becomes difficult to see above the Hawking--Page transition. The photon sphere approaches the event horizon as the black hole becomes small, and approaches the AdS boundary as the black hole becomes large. Note that the effective potential is shifted vertically upward, compared to the asymptotically flat case. }
    \label{fig:effective_potential_radial}
\end{figure}
\begin{equation}
    h'(\psp) = 0.
    \label{eq:pspformula}
\end{equation}
This gives the radial position of the photon sphere as:
\begin{equation}
    \psp =  \left(\frac{1}{(d-1) M}\right)^{\frac{1}{3-d}}.
\end{equation}
The mass of the black hole determines the location of the photon sphere in maximally symmetric spacetimes. This is because the effective potential can be written as:
\begin{equation}
    h(r) = \frac{1}{r^2} + \frac{k}{L^2} - \frac{2 M}{r^{d-1}} ,
\end{equation}
where $k = 0, 1, -1$ for asymptotically flat, AdS, and dS spacetimes, respectively. From here on, we choose $L=1$ for convenience and specialize to $d=4$. Since $k$ appears in a term without $r$ dependence, the location of the photon sphere \eqref{eq:pspformula} in terms of the mass is the same in all three types of spacetime.\footnote{The radius $R$ of an AdS-Schwarzschild black hole depends in a more complicated way on its mass: $M =  R^{d-1} \left(L^2+R^2\right)/2 L^2$.}  Then for $d=4$ we have: 
\begin{equation}
    \psp = 3 M.
\end{equation}
The effective potential is related to its counterpart in flat space by: 
\begin{equation}
h(r) - k = h_{\text{flat}}(r).
\end{equation}
Since $k$ does not depend on $r$, it follows immediately that the $n$-th derivative of $h(r)$ with $n\ge1$ satisfies
\begin{equation}
    h^{(n)}(r)\bigg|_{r = (2+n)M} = 0.
    \label{eq:maxaccel}
\end{equation}
For example, as a consequence of Equation \eqref{eq:maxaccel}, the photon sphere occurs at $r=3M$ and the maximum acceleration of the particle occurs at $r=4M$. 

For asymptotic AdS spacetime, the effective potential on the boundary is $h(\infty) = 1/L^2 = 1$, so to leave the boundary massless particles need $\omega>l$, or equivalently, $b > 1$. They will fall into the black hole if $b$ exceeds the maximum value of the effective potential, which is given by its value at the location of the photon sphere: 
\begin{equation}
h(\psp) = 1+\frac{1}{27 M^2}.
\label{eq:PSP_impact_parameter}
\end{equation}
In summary, for the geodesic to travel between two points on the boundary, its impact parameter $b = \omega/l$ must lie in the range: 
\begin{equation}
    1 \le \frac{\omega^2}{l^2} \le  1+\frac{1}{27 M^2} := b_{\text{crit}}^2.
    \label{eq:geodesiccross}
\end{equation}
Equivalently:
\begin{equation}
    h(2M) < b^2 < h(3M).
    \label{eq:impact_boundaries}
\end{equation}

It is interesting to consider the limit where the black hole becomes large ($M \rightarrow \infty$), which is known as the planar limit. The critical impact parameter will be the reciprocal of the AdS radius, and the effective potential becomes: 
\begin{equation}
    h_{\text{planar}}(r) = \frac{1}{r^2} - \frac{2M}{r},
\end{equation}
which does not have a photon sphere in the bulk spacetime. In this limit, the potential increasingly resembles the asymptotically flat case, but even in the limit as $M\rightarrow 0$, the potential is still shifted up by $1$. It can be seen from Equations \eqref{eq:PSP_impact_parameter}, \eqref{eq:geodesiccross}, and \eqref{eq:impact_boundaries} that the photon sphere approaches the boundary as $M \rightarrow \infty$, and that the impact parameter for geodesics traveling between boundary points lies in an increasingly narrow range. The opposite limit is also interesting. The photon sphere moves toward the event horizon when we take the limit where the black hole becomes small $(M \rightarrow 0)$. 

\section{The Scalar Wave Equation on a Schwarzschild Background}
\label{sec:KG}
Here we motivate our findings by exploring the well-known analogy, in the eikonal limit, between the wave equation for a minimally-coupled massless scalar field $\Phi(t,r,\theta,\phi)$ on an AdS$_4$-Schwarzschild background and the equation of motion for a null geodesic. The field must satisfy the Klein--Gordon equation: 
\begin{equation}
    g_{\mu \nu}\nabla^\mu \nabla^\nu \Phi(t,r,\theta,\phi) = 0.
     \label{eq:KG}
\end{equation}
Specializing to a Schwarzschild black hole with blackening function $F(r)$, this yields: 
\begin{equation}
        \frac{1}{F}\partial_t^2 \Phi + F \partial^2_r \Phi + \frac{\partial_r (r^2 F)}{r^2}\partial_r\Phi + \frac{\nabla^2}{r^2}\Phi = 0,
\end{equation}
where $\nabla^2$ is the scalar Laplacian on $S^{d-2}$. Since this problem has an azimuthal symmetry, we can solve this equation by expanding its solution in spherical harmonics $Y_{lm}$ with $m=0$:
\begin{equation}
    \Phi(t,r,\theta,\phi) = e^{- i \omega t} \sum_l c_l \phi_l(r) Y_{l\,0}(\theta).
\end{equation}
This approach is useful due to the property:
\begin{equation}
\nabla^2 Y_{lm}(\theta,\phi) = \frac{-l(l+1)}{r^2} Y_{l m}(\theta,\phi).
\end{equation}
This leads to the differential equation for the radial field with angular momentum $l$:
\begin{equation}
    F(r)^2 \phi_l''(r) + \left( \frac{2 F(r)^2}{r} + F(r) F'(r) \right) \phi_l'(r) + \left( \omega^2 - \frac{l(l+1)F(r)}{r^2}\right)\phi_l(r) = 0.
    \label{eq:uglyKG}
\end{equation}
\subsection{Tortoise Coordinates and Schr\"odinger Form}

The equations of motion \eqref{eq:uglyKG} simplify in tortoise coordinates:
\begin{equation}
    r_* := \int \frac{1}{F(r)} dr \Leftrightarrow \frac{d}{dr_*} := F(r) \frac{d}{dr},
    \label{eq:tortoise}
\end{equation}
because: 
\begin{align}
\label{eq:derivatives}
    \left(\frac{d}{dr_*}\right) \phi_l(r_*) &= F(r) \phi_l'(r),\\
    \left( \frac{d}{dr_*} \right)^2 \phi_l(r_*) &= F(r) F'(r) \phi_l'(r) + F(r)^2 \phi_l''(r).
\end{align}
It is straightforward to use the above equations to show that: 
\begin{equation}
\left( \frac{d}{dr_*} \right)^2 \phi_l(r_*) + \left( \frac{2 F(r)}{r}\right) \left(\frac{d}{dr_*}\right)\phi_l(r_*) + \left(\omega^2 - \frac{l(l+1)F(r)}{r^2}\right)\phi_l(r_*) = 0,
\label{eq:Helmholtzform}
\end{equation}
where the blackening function $F(r)$ is evaluated in the original $r$-coordinates. Now conveniently, this equation resembles the radial Helmholtz equation\footnote{The Helmholtz equation is obtained for flat space, i.e., $F(r) = 1$.}, which takes the following form for a radial field $R(r)$: 
\begin{equation}
    \left(\frac{d}{dr}\right)^2 R(r) + \frac{2}{r} \left(\frac{d}{dr}\right) R(r) + \left( k^2 - \frac{l(l+1)}{r^2}\right) R(r) = 0.
\end{equation}
This has an effective potential term $1/r^2$, and the term proportional to $l(l+1)$ is known as the ``centrifugal'' term. This gives a potential barrier that pushes solutions radially outward away from the origin, where the effective potential achieves its diverging global maximum. From this we can draw some intuition for the effect of the centrifugal term in Equation \eqref{eq:Helmholtzform} -- the effective potential is now $F(r)/r^2$ as in Equation \eqref{eq:effpot}, and the centrifugal term pushes solutions outward away from the photon sphere, which is the local maximum for the effective potential. For these reasons, we anticipate that solutions $\phi_l(r_*)$ corresponding to a source on the boundary will be found outside the photon sphere for large $l$. 

To convert Equation \eqref{eq:Helmholtzform} to the Schr\"odinger form, we make use of the field redefinition described in Appendix \ref{sec:AppendixA} to obtain:\footnote{Root finding methods can be used to obtain $r(r_*)$. See Section \ref{sec:numerics}.}
\begin{equation}
    \left( \frac{d}{dr_*} \right)^2\psi_l(r_*) + \left( \omega^2 - \underbrace{\frac{l(l+1)F(r)}{r^2} - \frac{F(r)F'(r)}{r}}_{V(r)}\right)\psi_l(r_*) = 0,
    \label{eq:Schrodingerform1}
\end{equation}
where $\psi_l(r) := r \phi_l(r)$, and where we have specialized to $d=4$. There  Klein--Gordon equation in Schr\"odinger form clearly resembles the equation for the geodesics of massless particles \eqref{EQ:geodesic_equation}. The  multiplicative coefficient of $\psi_l(r_*)$ in Equation \eqref{eq:Schrodingerform1} is positive when: 
\begin{equation}
    \frac{\omega^2}{l(l+1)} > \frac{F(r)}{r^2} + \frac{F'(r) F(r)}{r l(l+1)}.
\end{equation}
In the limit where the angular momentum, $l$, becomes large, we obtain: 
\begin{equation}
 \frac{\omega^2}{l(l+1)} > \frac{F(r)}{r^2} = h(r),
 \label{eq:KGcross}
\end{equation}
which is roughly the same as the condition for null geodesics to travel between boundary points in Equation \eqref{eq:geodesiccross}. Since $\psi$ takes the form of the Schr\"odinger equation, we will analyze its behavior by using the semi-classical WKB approximation -- i.e., the wave must tunnel through a potential with a local maximum at the photon sphere. The second term, $F'(r)F(r)/r$, acts as a ``reflective" potential barrier located on the boundary:\footnote{It is well approximated by an almost vertical wall, especially for large $l$, as shown in Figure \ref{fig:small_bh_tortoise}.}
\begin{equation}
\label{eq:reflect}
    \frac{F'(r)F(r)}{r} = \frac{r^{-2 d-1} \left(L^2+r^2\right) \left(R r^d-r R^d\right) \left((d-1) R^d \left(L^2+r^2\right)+2 R r^{d+1}\right)}{L^4 R^2}.
\end{equation}
In asymptotically flat spacetimes, $L\rightarrow \infty$ yields $F'(r)F(r)/r \rightarrow 0$ at the boundary. But for asymptotically AdS$_d$ spacetimes, it is simple to show that this diverges as $F(r)F'(r)/r \sim 2 r^2/L^4$, which gives basically an infinite potential barrier at the boundary. Nonetheless, once a UV cutoff is imposed, one might hope this term can be ignored for sufficiently large $l$. This approach is useful in the next section. 

\section{An Amplification Formula for Sources}
\label{sec:amplification}

According to the AdS/CFT correspondence, the sources and expectation values of operators in the boundary CFT can be obtained by first solving the classical wave equation in the bulk and then performing an asymptotic expansion near the boundary to determine the leading and subleading terms in the field. In this note, we exclusively focus on massless scalar fields in the bulk, corresponding to certain scalar operators in the boundary CFT. We anticipate that our argument would go through mostly unchanged for Maxwell and gravitational fields, but we leave that intriguing prospect for future work.

\subsection{Brief Review of the AdS/CFT Correspondence}
First, we briefly review the AdS/CFT correspondence and specialize to our case. For a scalar field of mass $m$ the conformal dimension $\Delta$ in an AdS$_{d}$ bulk can be determined from \cite{Gubser:1998bc}: 
\begin{equation}
    \Delta(\Delta -d+1) = m^2.
\end{equation}
In our case $d=4$ and $m=0$, so we have:
\begin{equation}
    \Delta_{-}=0,\quad \Delta_{+}=3.
\end{equation}
According to the AdS/CFT dictionary, the field $\Phi$ near the asymptotic boundary expanded in terms of $1/r$ can be written as \cite{Klebanov:1999tb,Hashimoto:2018okj}:
\begin{equation}
    \Phi(t, r, \theta, \phi) = J_\mathcal{O}(t, \theta, \phi) - \frac{1}{2r^2}\left( \partial_t^2 - \nabla^2\right)J_{\mathcal{O}}(t, \theta, \phi)  + \frac{\expval{\mathcal{O}(t, \theta, \phi)}}{r^3} + \dots,
    \label{eq:bulkoperatorcorrespondence}
\end{equation}
where $J_\mathcal{O}$ is the oscillating source (with frequency $\omega$) for the scalar operator $\mathcal{O}$ dual to $\Phi$ in the boundary CFT, so that $J_\mathcal{O}$ and the expectation value of $\mathcal{O}$ can be read off from the asymptotic expansion. The source is the boundary value of the field at the UV cutoff, while the response function $\expval{\mathcal{O}}$ corresponds to the coefficient of the $1/r^3$ term. 

\subsection{An Amplification Formula}

Here we derive a formula that precisely describes the amplification of fields $\phi(r_*)$ in the AdS-Schwarzschild spacetime. These are dual to sources $J_\mathcal{O}$, with frequency $\omega$, that drive operators $\mathcal{O}$ with angular momentum $l$ on the boundary CFT dual to the AdS-Schwarzschild bulk. Note that, in exact analogy to massless particles traveling along null geodesics, we are free to define an impact parameter for the pair $(J_\mathcal{O}(\omega),\mathcal{O}_l)$ as $b^2 := \left(\frac{\omega}{l}\right)^2$.\footnote{In the next section, we will see that the critical impact parameter, which separates massless particles which fall into the black hole from those which do not, equals the Lyapunov exponent; that is, $b_{\text{crit}} =\lambda$.} 

Consider the WKB approximation for the auxiliary field in the Schr\"odinger form \eqref{eq:Schrodingerform1}:
\begin{equation}
    \psi_l(r_*) \approx \frac{1}{\sqrt{\abs{p(r_*)}}}e^{i \int p(r_*) dr_*},
    \label{eq:psiapprox}
\end{equation}
where $p(r_*) = \sqrt{\omega^2 - V(r_*)}$  is the ``classical momentum'', with $V(r_*)$ given by Equations \eqref{eq:classical_momentum} and \eqref{eq:Schrodingerform1}. When $l$ is sufficiently large, the ``reflective" term at the AdS boundary drops out of the region of interest, and the geodesic condition \eqref{eq:geodesiccross} essentially determines whether $p(x)$ is real or imaginary -- when it is satisfied, $p(x)$ is real, and there will be no potential barriers for the wavefunction $\psi_l(r_*)$ to tunnel through and enter the region of interest. 

For modes analogous to particles falling in the black hole, the photon sphere is a classically allowed region. Then through the WKB approximation \eqref{eq:psiapprox}, the amplitude for the solution $\psi_l(r_*)$ to Equation \eqref{eq:Schrodingerform1} is well approximated by:
\begin{equation}
    |\psi_l(r_{*})| \approx \frac{A}{\sqrt[4]{\omega^2 - l(l+1)\left(\frac{f(r)}{r^2}\right)}}.
    \label{eq:amplitudeapprox1}
\end{equation}
Note that $r = r(r_*)$. Here $A$ is a constant that must be determined. For an asymptotically AdS boundary, the effective potential $h(r) \equiv \frac{f(r)}{r^2} \rightarrow 1$. This gives the condition: 
\begin{equation}
    |\psi_l(r_{*B})| \approx \frac{A}{\sqrt[4]{\omega^2 - l(l+1)}}.
    \label{eq:amplitudeapprox2}
\end{equation}
Just as in non-relativistic quantum mechanics, where $A$ would be determined through normalization, the coefficient $A$ must be found through physical considerations. As explained in Appendix \ref{sec:AppendixA}, the bulk field $\phi(r_*)$ is related to $\psi(r_*)$ through $r(r_*) \phi(r_*) = \psi(r_*)$. 

At the boundary we have:
\begin{equation}
    \frac{A}{\left(\omega^2 - l(l+1)\right)^{1/4}} \approx |\psi_l(r_{*B})|  = r_B |\phi(r_{*B})|,
\end{equation}
where $r_{*B}$ is the tortoise coordinate at the AdS boundary. This is interesting because it suggests a natural value for the UV cutoff $r_B$ in the AdS/CFT correspondence. It would be interesting to understand this in greater detail. Working along this line,  we obtain a formula that describes the amplification of the field. The general solution for $\phi$ is asymptotically:
\begin{equation}
    \phi(r_*)=B e^{i p r_*}+C e^{-i p r_*}.
\end{equation}
Here $p$ is the classical momentum: $p=\sqrt{\omega^2 - l^2\frac{f(r)}{r^2}}$. Choosing the outgoing mode, and performing a Taylor expansion near the AdS boundary (where $r_* = 0$), gives: 
\begin{equation}
    \frac{A}{r_B\left(\omega^2 - l(l+1)\right)^{1/4}} \approx\left\vert \frac{\psi(r_{*B})}{r_B} \right\vert = |\phi(r_{*B})| \approx p_B r_{*B} \approx \frac{p_B}{r_B} = \frac{ \sqrt{\omega^2 - l(l+1)}}{r_B}.
\end{equation}
Just one normalization constant is needed for the fields, due to Equation \eqref{eq:psiandphi}, so we have set $B=1$. We also used Equation \eqref{eq:tortoise_Expansion}. To have the right scaling behavior on both sides of this equation, we must choose:
\begin{equation}
    A = \left( \omega^2 - l(l+1)\right)^{3/4}.
    \label{eq:normalizationofA}
\end{equation}
Alternatively, the boundary value of $\psi$ can be found numerically to determine $A$. This is done by first solving the differential equation for $\phi(r_*)$ numerically, normalizing the boundary value of $\phi(r_*)$ to $1$\footnote{Here we follow the convention in \cite{Hashimoto:2018okj}, which normalizes the boundary value of $\phi(r_*)$ to unity.}, and then determining $\psi(r_*)$ using Equation \eqref{eq:psiandphi}.  The numerical solution for $\psi(r_*)$, when determined in this way, is found to diverge beyond the UV cutoff suggested above. The boundary value of $\psi$ is then found to be:
\begin{equation}
     |\psi_l(r_{*B})| \approx \sqrt{\omega^2 - l(l+1)},
\end{equation}
which then determines $A = \left( \omega^2 - l(l+1)\right)^{3/4}.$ The upshot is that the analytic and numerical approaches give the same answer. The WKB formula for $|\psi(r_*)|$ is then:
\begin{equation}
    |\psi_l(r_*)| \approx \frac{\left(\omega^2 - l(l+1)\right)^{\frac{3}{4}}}{\sqrt[4]{\omega^2 - l(l+1)\frac{f(r)}{r^2}}},
\end{equation}
where $r = r(r_*)$ is the \textit{original} coordinate in terms of the Tortoise coordinate. Finally, using Equation \eqref{eq:psiandphi}, we obtain a formula describing the amplification of the bulk field $\phi(r_*)$: 
\begin{equation}
    \frac{|\phi_l(r_*)|}{|\phi_l(r_{*B})|} \approx \frac{\left(\omega^2 - l(l+1)\right)^{\frac{3}{4}}}{r(r_*) \sqrt[4]{\omega^2 - l(l+1)\frac{f(r)}{r^2}}}.
    \label{eq:amplification_formula}
\end{equation}

The field is amplified by two distinct physical effects. First there is the factor of $r^{-1}$, which takes its maximum value at the horizon of the black hole. There is also $\omega^2 - l(l+1) \frac{f(r)}{r^2}$, which describes the amplification of the field at the photon sphere. The latter diverges as the critical impact parameter is approached. Note that the amplitude of the field $\psi$, and the effective potential $h(r) \equiv \frac{f(r)}{r^2}$, take their maximum value at the photon sphere: 
\begin{equation}
    |\phi_l(r_{*PS})|  \approx \frac{\left(\omega^2 - l(l+1)\right)^{\frac{3}{4}}}{r_{PS} \sqrt[4]{\omega^2 - l(l+1) \lambda^2}}.
\end{equation}
Here we have used Equation \eqref{eq:geodesiccross}; $r_{*PS}$ and $r_{PS}$ are the locations of the photon sphere in tortoise coordinates and regular coordinates, respectively. Note that $\phi$ does not necessarily achieve its maximum value at the photon sphere, even though $\psi$ does. Since $r_{PS} = 3M$, the height of this peak is more intense for small AdS black holes. The peak is completely washed away in the planar AdS limit: $R \rightarrow \infty$. 

This formula is found to precisely match our numerical results; see Figures \ref{fig:HP_amplitudes}--\ref{fig:tiny_amplitudes}. For both $\phi(r_*)$ and $\psi(r_*)$, the amplitude at the photon sphere diverges when the potential barrier at the photon sphere becomes classically forbidden. Beyond this point the amplification formula breaks down because the amplitude will decay within the photon sphere, which signals the existence of quasinormal modes.\footnote{In an earlier version of this note, we used this argument to suggest the existence of quasinormal modes. We had not known this was pointed out much earlier by Festuccia and Liu \cite{Festuccia:2008zx}. We thank Matthew Dodelson and Cristoforo Iossa for bringing this to our attention.} In the next section, we obtain a simple formula for the first few overtones by approximating the potential at the photon sphere as an inverted harmonic oscillator.  Note that this effect is more intense for small black holes, which are below the Hawking-Page phase transition \cite{Hawking:1982dh,Witten:1998zw}, and that this effect is suppressed for large black holes. The suppression occurs because the photon sphere aligns with the AdS boundary in the limit as $R\rightarrow1$, and in this limit $\lambda \rightarrow 1$.\footnote{It is interesting to note that the AdS boundary satisfies the conditions for a photon sphere, i.e. $h'(r_B)=0$.} Note the divergence beyond the natural value of the cutoff, discussed above; it can be removed by choosing that value for the cutoff. 
\section{The Lyapunov Exponent, the Effective Potential, and QNMs}
\label{sec:Lyapunov}

The Lyapunov exponent for QNMs in asymptotically flat backgrounds was determined in \cite{Cardoso:2008bp}. It plays a fundamental role in determining the QNMs of perturbed black holes in asymptotically flat spacetime: 
\begin{equation}
\lambda = \frac{1}{\sqrt{2}} \sqrt{-\frac{\psp^2}{F(\psp)}\left(\frac{d^2}{d r_*^2} \frac{F(r)}{r^2}\right)\bigg|_{r=\psp}}\,.
\label{eq:Lyapunov}
\end{equation}
We will see that a Taylor series expansion at the photon sphere gives an inverted harmonic oscillator depending on the Lyapunov exponent. 

The main contribution to the effective potential at large $l$ comes from the centrifugal term proportional to $l(l+1)$ in \eqref{eq:Schrodingerform1}. Since $h'(\psp)=0$, its Taylor series is:
\begin{align}
    \frac{V(r_*)}{l(l+1)} &= h(\psp) + \frac{1}{2} \left( \frac{d}{dr_*} \right)^2 h(r) \bigg|_{\psp} (r_*-\psp)^2\\
    &= h(\psp) \left( 1 - \lambda^2 (r_*-\psp)^2 \right).
    \label{eq:Lyapunov_approx}
\end{align}
This is an inverted harmonic oscillator potential, a potential barrier for the wavefunction $\psi_l$ to cross. From the roots of the quadratic polynomial, the width of the inverted harmonic oscillator potential is
\begin{equation}
    \delta =\frac{2}{\lambda},
    \label{eq:width}
\end{equation}
which provides a useful estimate for the width of the actual potential barrier. For a sufficiently large angular momentum $l$, we can estimate the decay percentage of $\psi_l$ using the WKB approximation. 

\begin{figure}

    \centering
    \includegraphics[width=\linewidth]{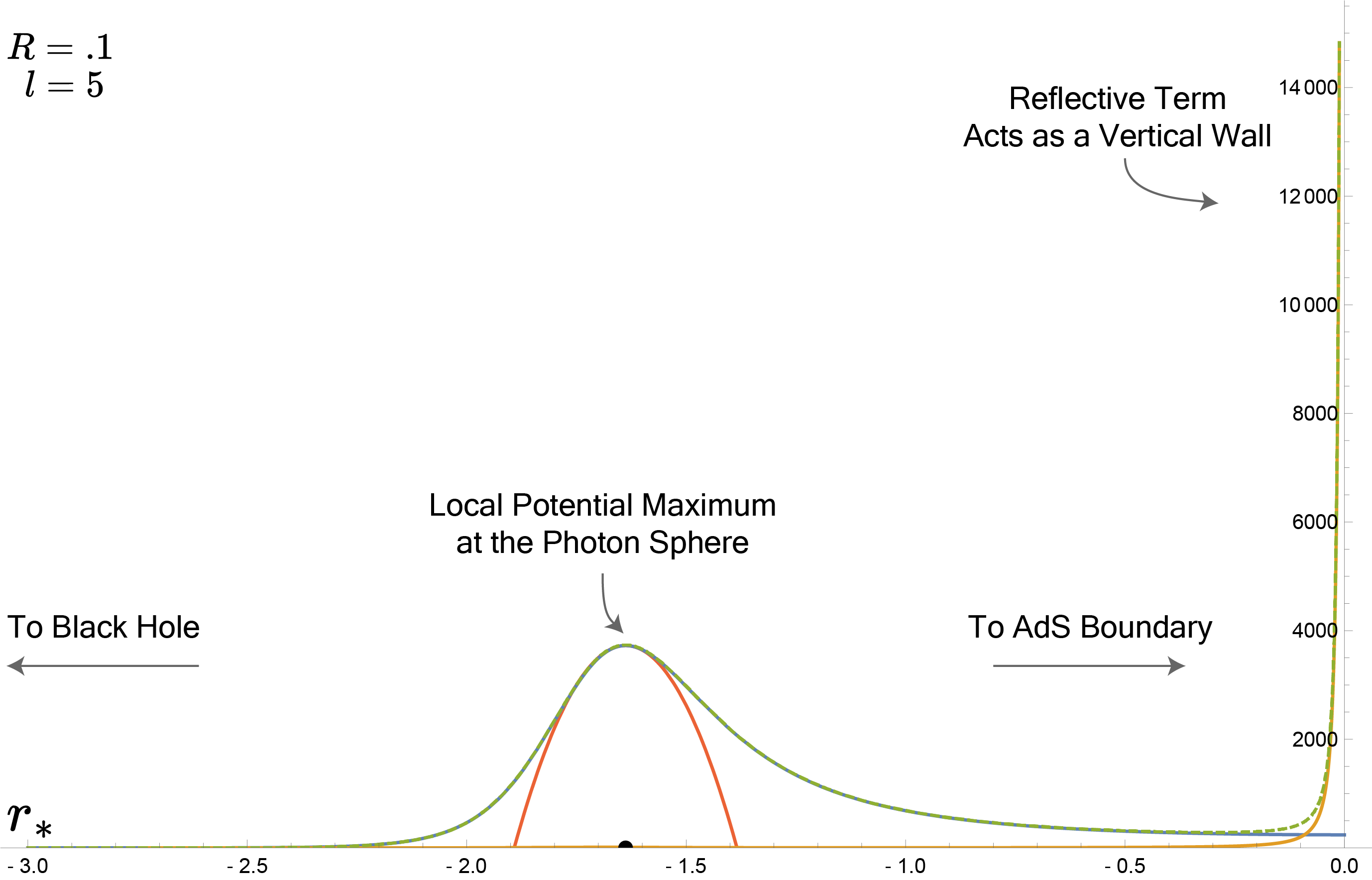}
    \caption{Here we show $V(r)$ for the auxiliary $\psi_l$ field in tortoise coordinates. The green curve is the actual value of the effective potential, the blue curve (which mostly overlaps with the green curve) is the centrifugal term $l(l+1)F(r)/r^2$, and the orange curve is the reflective term $F(r)F'(r)/r$. As a general rule, the blue curve approximates the effective potential  with high accuracy when $l$ is large. The red curve approximates the effective potential near the photon sphere in terms of the Lyapunov exponent \eqref{eq:Lyapunov_approx}. It is quite accurate for black holes smaller than the order of the Hawking--Page phase transition $(R=1)$. Since the ``reflective" boundary potential is approximately a vertical wall, especially at large $l$, it should be possible to study this system using a WKB-style analysis. The photon sphere lies at approximately $r_* = - \pi/2$, as predicted by Equation \eqref{eq:pi2}.} 
    \label{fig:small_bh_tortoise}
\end{figure}

The reflective term \eqref{eq:reflect} acts as a hard wall at large $l$, as shown in Figure \ref{fig:small_bh_tortoise}. There is a potential barrier at the photon sphere, and a field driven by a source of  frequency $\omega$ can tunnel through that barrier if: 
\begin{equation}
    \frac{\omega^2}{l(l+1)} < h(\psp) \left( 1 - \lambda^2 (r-\psp)^2 \right).
\end{equation}
The next step is to compute the Lyapunov exponent: 
\begin{equation}
\lambda^2 = -\frac{1}{2} \frac{\psp^2}{F(\psp)}\left(\frac{d^2}{d r_*^2} \frac{F(r)}{r^2}\right)\bigg|_{r=\psp} = -\frac{1}{2}\left(\frac{1}{1 + \frac{1}{27M^2}}\right)\left( F(r) \frac{d}{dr} \right)\left( F(r) \frac{d}{dr} \right)\left( \frac{F(r)}{r^2}\right)\bigg|_{r=\psp}. 
\end{equation}
Surprisingly, this yields the critical impact parameter \eqref{eq:geodesiccross}: 
\begin{equation}
    \lambda^2 = 1 + \frac{1}{27M^2} = h(\psp) = b_{\text{crit}}^2
    \label{eq:Lyapunov_Surprise}
\end{equation}
The upshot is that the Lyapunov exponent is the critical value of the impact parameter that distinguishes null geodesics that fall into the event horizon from those that do not. Furthermore, the width of the potential barrier follows from Equation \eqref{eq:width}:
\begin{equation}
    \delta = \frac{2}{\sqrt{1 + \frac{1}{27M^2}}}.
\end{equation}
This estimation for the effective potential is highly accurate for large $l$, even near the asymptotic boundary. It turns out that the reflective term, $F(r)F'(r)/r$, also features a small peak at the photon sphere, but this can be neglected for sufficiently large $l$. For the most part, the reflective term's peak at the photon sphere matters only for black holes with very small masses. The effectiveness of our approximation is illustrated in Figure \ref{fig:small_bh_tortoise}.

\subsection{The Lyapunov exponent and AdS-Schwarzschild Black Holes}

The simple form of the Lyapunov exponent makes it possible to rephrase many key quantities in terms of it. We begin by solving for the mass of the black hole: 
\begin{equation}
    M = \frac{1}{\sqrt{27 \lambda ^2-27}}.
\end{equation}
In Schwarzschild coordinates, we use the known equation for the mass in terms of the black hole radius $R$ to solve for the mass in terms of the Lyapunov exponent:
\begin{equation}
    M = \frac{R(R^2+1)}{2},
\end{equation}
so that
\begin{equation}
    R = \frac{\sqrt[3]{\lambda +1}-\sqrt[3]{\lambda -1}}{\sqrt{3} \sqrt[6]{\lambda ^2-1}}.
\end{equation}
The Hawking--Page transition occurs when the radius of the black hole drops below the AdS radius -- i.e., $R=1$. Equivalently, it occurs when $M=1$. Then the Hawking--Page transition, in terms of the Lyapunov exponent, takes place when:
\begin{equation}
    \lambda = \sqrt{\frac{28}{27}}.
\end{equation}
Meanwhile, the temperature of the Schwarzschild black hole is: 
\begin{equation}
    T = \frac{2}{F'(r)}\bigg|_{r=R}=\frac{2 R}{1 + 3 R^2} = \frac{2 \left(\sqrt[3]{\lambda +1}-\sqrt[3]{\lambda -1}\right)}{\sqrt{3} \sqrt[6]{\lambda ^2-1} \left(\sqrt[3]{\frac{\lambda +1}{\lambda -1}}+\sqrt[3]{\frac{\lambda -1}{\lambda +1}}-1\right)}.
\end{equation}
This achieves its maximum value at what is known as the spinodal point, which turns out to be related to the golden ratio $\phi$\footnote{Another curious appearance of the golden ratio in black holes in a different background metric can be found in \cite{Cruz:2017rbk}.}: 
\begin{equation}
    \lambda = \frac{\sqrt{5}}{2} = \phi -1/2.
\end{equation}
It would be amusing to find out if this is related to the curious numerological fact that, for null geodesics with maximum radial acceleration in the \emph{de Sitter} spacetime, the two classical turning points occur in the golden ratio \cite{Cruz:2017rbk}. 

With some effort, it is possible to compute an analytic formula for the location of the photon sphere in tortoise coordinates in terms of the Lyapunov exponent. Its location is: 
\begin{equation}
P(\lambda) := r_{*,\text{ps}}=\frac{\sqrt[6]{\lambda +1}}{2 \sqrt{6} \sqrt[6]{\lambda-1}} \bigg[ A(\lambda) + B(\lambda) + C(\lambda) \bigg],
\label{eq:photonsphere_location}
\end{equation}
where functions $A(\lambda)$, $B(\lambda)$, and $C(\lambda)$ are:
\begin{equation}
\begin{aligned}
A(\lambda)&=-\frac{2 \sqrt[3]{(\lambda -1)^2 (\lambda +1)} \left(\sqrt[3]{\frac{\lambda -1}{\lambda +1}}-1\right) \log \left(\frac{\sqrt[3]{(\lambda -1)^2 (\lambda +1)}-\sqrt[3]{(\lambda -1) (\lambda +1)^2}+1}{\sqrt{3} \sqrt{\lambda ^2-1}}\right)}{-\sqrt[3]{(\lambda -1)^2 (\lambda +1)}+\sqrt[3]{(\lambda -1) (\lambda +1)^2}+\lambda -1},\\
    B(\lambda)&=\frac{\left(\left(1+i \sqrt{3}\right) \sqrt[3]{\frac{\lambda -1}{\lambda +1}}+i \sqrt{3}-1\right) \log \left(\frac{1}{6} \left(\frac{\left(\sqrt{3}-3 i\right) \sqrt[3]{\lambda +1}}{\sqrt[6]{\lambda ^2-1}}-\frac{\sqrt{3} \left(1+i \sqrt{3}\right) \sqrt[6]{\lambda ^2-1}}{\sqrt[3]{\lambda +1}}+\frac{2 \sqrt{3}}{\sqrt{\lambda ^2-1}}\right)\right)}{1-\frac{\left(-i \sqrt[3]{\frac{\lambda -1}{\lambda +1}}+\sqrt{3} \sqrt[3]{\frac{\lambda -1}{\lambda +1}}+i+\sqrt{3}\right)^2}{4 \sqrt[3]{\frac{\lambda -1}{\lambda +1}}}},\\
    C(\lambda)&=\frac{\left(\left(1-i \sqrt{3}\right) \sqrt[3]{\frac{\lambda -1}{\lambda +1}}-i \sqrt{3}-1\right) \log \left(\frac{1}{6} \left(\frac{\left(\sqrt{3}+3 i\right) \sqrt[3]{\lambda +1}}{\sqrt[6]{\lambda ^2-1}}+\frac{i \sqrt{3} \left(\sqrt{3}+i\right) \sqrt[6]{\lambda ^2-1}}{\sqrt[3]{\lambda +1}}+\frac{2 \sqrt{3}}{\sqrt{\lambda ^2-1}}\right)\right)}{1-\frac{\left(i \sqrt[3]{\frac{\lambda -1}{\lambda +1}}+\sqrt{3} \sqrt[3]{\frac{\lambda -1}{\lambda +1}}-i+\sqrt{3}\right)^2}{4 \sqrt[3]{\frac{\lambda -1}{\lambda +1}}}}.\\
\end{aligned}
\end{equation}
It is possible to use \textit{Mathematica} to compute the limit of Equation \eqref{eq:photonsphere_location} as $\lambda \rightarrow \infty$, which corresponds to $m \rightarrow 0$. Remarkably, it turns out to be:
\begin{equation}
    \lim_{\lambda \rightarrow \infty}r_{*,\text{ps}} = \frac{\pi}{2}.
    \label{eq:pi2}
\end{equation}
\subsection{Approximate Large-$l$ QNMs for AdS-Schwarzschild}
Provided the mass of the black hole is small enough -- in particular, below the Hawking--Page transition -- we can write our WKB potential as in Figure \ref{fig:small_bh_tortoise}. Writing everything in terms of the Lyapunov exponent gives: 
\begin{equation}
    \phi_l ''(r_*) +  \left(\omega ^2-\lambda ^2 l (l+1) \left(1-\lambda ^2 (r_*-P(\lambda))^2\right)\right)\phi_l (r_*)=0
    \label{eq:lyapunov_potential}
\end{equation}
We can use this to obtain the QNMs of the system, i.e., the modes corresponding to a vanishing source on the boundary. This differential equation admits closed-form solutions in terms of parabolic cylinder functions $D_\nu(z)$. Cumbersome solutions can be obtained using \textit{Mathematica}, where $P(\lambda$) is given by \eqref{eq:photonsphere_location}. The QNMs can probably be found numerically from those closed-form solutions. It is simpler to proceed by approximation. An analogous scenario involving the parabolic cylinder functions is described in Section 4.2 of \cite{Berti:2009kk}, where they explain that the asymptotic behavior of cylinder functions requires that $\nu$ is an integer.\footnote{An analogous argument also appears in \cite{Cardoso:2008bp}.} For a differential equation of the form:
\begin{equation}
\frac{d^2 \Psi}{d r_*^2}+\left[Q_0+\frac{1}{2} Q_0^{\prime \prime}\left(r_*-\bar{r}_*\right)^2\right] \Psi=0,
\end{equation}
where $Q_0$ is the coefficient of the second term in the Taylor series of \eqref{eq:lyapunov_potential} at the local maximum (in our case, the Lyapunov exponent \eqref{eq:Lyapunov}), the condition for $\nu$ to be integer is:
\begin{equation}
    \frac{Q_0}{\sqrt{2 Q_0''}} = i \left(n + \frac{1}{2}\right).
\end{equation}
We can rewrite our differential equation as follows:
\begin{equation}
    \phi_l''(r_*) + \underbrace{\left(\omega^2-\lambda ^2 l (l+1)\right)}_{Q_0} +  \underbrace{l (l+1)\lambda ^4 }_{\frac{1}{2}Q_0''}\left(r - P(\lambda)\right)^2 \phi_l(r_*) = 0
\end{equation}
So for sufficiently large $l$, the QNMs are given by, for $n \in (1, 2, 3, ...)$:
\begin{equation}
    \frac{\omega ^2-\lambda ^2 l (l+1)}{\sqrt{4 \lambda ^4 l (l+1)}} = i \left(n - \frac{1}{2}\right),
\end{equation}
\begin{equation}
    \implies \omega = \lambda  \sqrt[4]{l (l+1)} \sqrt{\sqrt{l (l+1)}+i(2n+1)}
    \label{eq:QNMs}
\end{equation}

The first QNM becomes available at roughly $\omega = \lambda l$, which matches the transition between particles falling into the black hole (no QNMs) to those which do not (the QNMs). The real part of the mode turns out to be roughly proportional to $l$, while the imaginary part is roughly evenly spaced in $n$.\footnote{This approximation is less accurate at large $n$. During the preparation of this note, \cite{Hashimoto:2023buz} appeared, which gives a more accurate answer at large $n$. Their approach is made more precise, but also more complicated, by the ``smoothed-out" inverted potential well known as the Posch-Teller potential.} Since the local maximum at the photon sphere roughly coincides with the boundary for black holes of the order of the Hawking--Page transition, we suspect this analysis applies only to small black holes. Note the proportionality to the Lyapunov exponent \eqref{eq:Lyapunov_Surprise}, which increases rapidly as the mass of the black hole drops below the Hawking--Page phase transition at $M=1$. This is consistent with the rapid growth of the peak, at the photon sphere, in the effective potential pictured in Figure \ref{fig:small_bh_tortoise}. 

\section{Numerical Evidence for the Amplification Formula}
\label{sec:numerics}
Here we present our numerical evidence for the amplification formula \eqref{eq:amplification_formula}. Our findings provide strong evidence for the formula's accuracy, though it has some limitations related to the breakdown of the WKB approximation under certain conditions. In this section, we first introduce our numerical procedure and  present our numerical results. Then we discuss the conditions under which the approximation fails. 

\subsection{Numerical Procedure}
\label{sec:divide}
We evaluated the fields $\psi_l(r_*)$ and $\phi_l(r_*)$, which satisfy Equations \eqref{eq:Schrodingerform1} and \eqref{eq:Helmholtzform} respectively, using a setup similar to the one outlined in \cite{Hashimoto:2018okj}. Recall that the field $\psi_l(r_*)$ satisfies the differential equation:
\begin{equation}
    \left( \frac{d}{dr_*} \right)^2\psi_l(r_*) + \left( \omega^2 - \frac{l(l+1)F(r)}{r^2} - \frac{F(r)F'(r)}{r}\right)\psi_l(r_*) = 0.
    \label{eq:Schrodingerform4}
\end{equation}
Here $r_*$ is the tortoise coordinate \eqref{eq:tortoise} given by $r_* = \int dr/F(r)$. As we have seen, one can determine the tortoise coordinate analytically and phrase it in terms of the Lyapunov exponent $\lambda$. While those expressions are unwieldy, the asymptotic expansions are simple: 
\begin{equation}
r_*=-\frac{1}{r}+\frac{1}{3 r^3}+\mathcal{O}\left(\frac{1}{r^4}\right), \quad r=-\frac{1}{r_*}+\frac{r_*}{3}+\mathcal{O}\left(r_*^2\right).
\label{eq:tortoise_Expansion}
\end{equation}
We used \textit{Mathematica} to invert the tortoise coordinate for $r(r_*)$, using \textsf{FindRoot}, and then solved Equation \eqref{eq:Helmholtzform} numerically. Because the horizon lies at $r_* = -\infty$ in tortoise coordinates and the AdS boundary lies at $r=\infty$ in regular coordinates, from \eqref{eq:tortoise_Expansion} we have
\begin{equation}
\label{eq:rule}
\begin{aligned}
    r_*(r=R) &= - \frac{1}{R},\\
    r_*(r = \infty) &= \epsilon.
    \end{aligned}
\end{equation}
The boundary conditions for $\phi_l$ at the horizon $(H)$ is that the wave must be in-going. Without loss of generality, we choose our boundary $(B)$ value for the field as:
\begin{equation}
\begin{aligned}
    \frac{d}{dr_*}\phi_l(r_*)\bigg|_H &= - i \omega,\\
    \phi_l(r_*)\bigg|_B &= 1.
    \label{eq:alternative}
    \end{aligned}
\end{equation}
To achieve the desired accuracy in our numerics, we decided against imposing Equation \eqref{eq:alternative} directly in \textsf{NDSolve}. Instead, we set $\phi(r_*)\big|_H = 1$ and then divided by the boundary value of the field to enforce $\eqref{eq:alternative}$. This grants us the freedom to choose our UV cutoff as $\epsilon = 10^{-5}$ and the location of the black hole as $r_* = -5$, which was accurate enough for our purposes. After solving \eqref{eq:Helmholtzform} we were left with an equation for $\phi_l(r_*)$, which is related to $\psi_l(r_*)$ by Equation \eqref{eq:psiandphi}; the inverse tortoise coordinate was then used to determine $\psi_l(r_*)$. 

\subsection{Numerical Results}

To verify the validity of the approximations leading to Equation \eqref{eq:amplification_formula}, we numerically computed the solutions to the differential equation \eqref{eq:Helmholtzform} for $\phi(r_*)$, and then obtained $\psi(r_*)$ -- which solves \eqref{eq:Schrodingerform1} -- using $\eqref{eq:psiandphi}$. These describe bulk fields in equivalent Schr\"odinger $\psi(r_*)$ and Helmholtz-like $\phi(r_*)$ perspectives, up to a field redefinition $\psi_l = r(r_*) \phi_l$, for various black hole masses where $\omega$ is the frequency of the driving source on the boundary.\footnote{The analysis at other frequencies proceeds identically.} 

The amplification formula \eqref{eq:amplification_formula} holds with a high degree of precision. Additionally, two distinct phase transitions, corresponding to the upper and lower limits of \eqref{eq:geodesiccross}, can be seen in our numerical results as the angular momentum $l$ of the dual operator $\mathcal{O}$ is varied. See Figures \ref{fig:HP_amplitudes}--\ref{fig:tiny_amplitudes}. The approximation is effective when the region inside the photon sphere is not classically forbidden for the $\psi(r_*)$ differential equation. The amplification formula \eqref{eq:amplification_formula} is violated as the critical impact parameter \eqref{eq:geodesiccross}, $b_c \equiv \frac{\omega}{l_c} = \lambda$, is approached.\footnote{The oscillating modes of $\phi_l(r_*)$ are divided about equally at the photon sphere near the impact parameter. This reminds us of a relatively recent paper \cite{Bardeen:2018omt}, in which it was argued that Hawking radiation is primarily generated near the photon sphere.}  Massless particles in the neighborhood of $b_c$ will orbit the photon sphere repeatedly before falling into the black hole or returning to the boundary, as shown in Figure \ref{fig:geodesics}. In precisely the same way, the amplification formula predicts a sharp peak in the amplitude of the field at the photon sphere near the critical impact parameter. Indeed, this makes physical sense because the WKB approximation predicts large amplitudes in regions where classical particles spend more time. The height \eqref{eq:Lyapunov_Surprise} and location \eqref{eq:pi2} of the peak depend on the mass of the black hole, with black holes well below the Hawking--being very sharply peaked. There is an additional effect where the field is suppressed to zero amplitude in the limit as $l \rightarrow \omega$, which is physically the same as a particle with large angular momentum being trapped on the AdS boundary. This can be understood physically; the AdS boundary satisfies the conditions for a photon sphere -- namely, $h'(r_B)=0$ -- and the interior of AdS becomes classically forbidden for $l > \omega$. In the planar black hole limit $R\rightarrow \infty$, these two transitions coincide as $\lambda \rightarrow 1$. 

\section{Conclusion}
\label{sec:conclusion}

It is well-known that boundary CFT operators $\mathcal{O}$ are holographically dual to bulk fields $\phi$ \eqref{eq:bulkoperatorcorrespondence} through the AdS/CFT correspondence \cite{Maldacena:1997re,Gubser:1998bc,Witten:1998qj}. Here we explored the connection between the Klein--Gordon and Schr\"odinger equations and derived an ``amplification formula" \eqref{eq:amplification_formula} that accurately predicts the amplitude of the field $\phi$ in the bulk theory. This discussion focused on implications for the AdS / CFT correspondence and was phrased in terms of the impact parameter $b$ for an operator $\mathcal{O}$, defined as the angular momentum of the operator, $l$, divided by the frequency of its driving source $\omega$. Depending on the value of their impact parameter, massless particles with energy $\omega$ and angular momentum $l$ leave the boundary, scatter off the photon sphere, or descend into the AdS-Schwarzschild black hole. See Figure \ref{fig:geodesics}. The upshot of this article is that bulk fields and their dual CFT operators (Figure \ref{fig:small_amplitudes}) are sensitive to this transition. There are immediate consequences for the ``photon sphere image" conjecture \cite{Hashimoto:2018okj,Hashimoto:2019jmw}, as we will show in an upcoming note \cite{Riojas:2023pje}.\footnote{The upshot is that the conjecture is very accurate when the long-lived modes, first identified in \cite{Festuccia:2008zx}, are not included in the sum. They were studied in \cite{Guo:2021bcw} for flat space.}

Certain qualitative changes in behavior for the scalar wave equation \eqref{eq:KG} are in exact analogy to the behavior of null geodesics in the bulk \eqref{EQ:geodesic_equation}. The transition occurs at the critical impact parameter \eqref{eq:geodesiccross}, as illustrated in Figure \ref{fig:geodesics}. This can be understood as follows. In Section \ref{sec:amplification}, we saw that an auxiliary radial field, $\psi(r_*) = r(r_*) \phi(r_*)$, is affected by a potential barrier at the photon sphere in the WKB approximation. Here $r_*$ is the tortoise coordinate \eqref{eq:tortoise}. For positive ``classical momentum" $p(x)$, the amplification formula \eqref{eq:amplification_formula} is accurate because the photon sphere is not classically forbidden, and the amplitude is not attenuated by tunneling effects. Massless particles with this impact parameter will fall in the black hole. 

It is well known that the Klein--Gordon equation is well approximated by geodesics in the eikonal limit.\footnote{There are many discussions of this fact; see, e.g., \cite{Berti:2009kk} and \cite{Hashimoto:2018okj, Hashimoto:2019jmw}. This limit is also useful in studying photon sphere signature in thermal two-point functions \cite{Dodelson:2023nnr}. For use in Kerr--Newman black holes, see \cite{Li:2021zct}.} Tunneling effects begin to play a role once the angular momentum of the operator becomes large enough. The first change occurs when the ``classical momentum" $p(x)$ from \eqref{eq:classical_momentum} becomes negative at the critical value $b_{\text{crit}}$ of the impact parameter.  Classical massless particles orbit the photon sphere for $l \gtrsim \omega/b_{\text{crit}}$. The bulk field $\phi(r_*)$ is not quite trapped outside the photon sphere; since $\psi(r_*)$ can tunnel through a potential barrier at the photon sphere, so can $\phi(r_*)$. In Section \ref{sec:Lyapunov}, we showed the critical impact parameter equals the Lyapunov exponent $\lambda$ of null geodesics orbiting the photon sphere, and that the first QNM occurs at $\omega \sim \lambda l$.  Near the critical value for the impact parameter, null geodesics repeatedly circle the photon sphere, before either returning to the asymptotic boundary or plunging into the black hole. In the same way, the field $\phi(r_*)$ becomes singular at this point, and QNMs activate.\footnote{We were not the first to point out that QNMs are connected to the orbits of massless particles \cite{Festuccia:2008zx}. We thank Matthew Dodelson and Christoforo Iossa for bringing this to our attention.}  There is an additional transition occurring at $l>\omega$ which is not conditional on the mass of the black hole. In this case the field $\phi(r_*)$ becomes trapped on the asymptotic boundary. These phenomena can be seen in Figures \ref{fig:HP_amplitudes}--\ref{fig:tiny_amplitudes}, in analogy with Figure \ref{fig:geodesics}.

We have found an intriguing relationship between the impact parameter $b = \omega/l$ of an operator $\mathcal{O}$ -- where $\omega$ is the frequency of the driving source and $l$ is the angular momentum of the operator -- and the region probed by its dual bulk field $\phi$.\footnote{David Berenstein, Ziyi Li, and Joan Sim\'on \cite{Berenstein:2020vlp} outlined a similar approach, and also highlighted the importance of the angular momentum of boundary operators, in an earlier note. We were unaware of their work until after this article was posted. One significant difference between their work and ours is that we obtained the amplification formula, performed the numerics, went farther with the QNMs, and showed $b_{\text{crit}} = \lambda$. As we were writing this article, our goal was to understand what we were seeing while performing the numerics of \cite{Hashimoto:2018okj,Hashimoto:2019jmw}, and we missed these works on ISCOs during our literature search.} It would be interesting to explore this connection further. The original motivation for this work was to build upon the ``photon sphere image conjecture" \cite{Hashimoto:2018okj,Hashimoto:2019jmw}, which attempts to reconstruct the image of an AdS photon sphere using the response function. We have taken a first step toward achieving this in \cite{Riojas:2023pje}, where we show the conjecture \textit{holds} -- this represents a significant improvement over previous work -- when the \textit{orbiting} modes are excluded from the sum. Our findings may be of interest to those studying QNMs in AdS/CFT, where the application of eikonal and WKB-inspired approximations has proved challenging, as noted in \cite{Cardoso:2008bp}. One might also explore the analytic solutions to \eqref{eq:lyapunov_potential}, or straightforwardly extend our analysis to massive and higher-spin fields. It would also be interesting to improve the precision of our analysis for the QNMs\footnote{Progress has already been made in \cite{Hashimoto:2023buz}, which appeared while this note was being prepared. It used the more precise Posch-Teller potential.}, or to understand if the impact parameter of operators is connected to the Hawking-Page phase transition \cite{Hawking:1982dh,Witten:1998zw}. Finally, it is well-known that Hawking radiation is highly sensitive to the angular momentum of its modes, and that this can be encoded in so-called greybody factors. Most of the radiation that probes the exterior region is s-wave radiation -- in other words, it is above the critical impact parameter\footnote{We used the convention where $b = \omega/l$. The other convention is $b=l/\omega$, which would place these modes below the critical impact parameter.} -- but the an observer inside the photon sphere would still be sensitive to the higher angular momentum modes.\footnote{These are conceptually related to the modes that we excluded from the sum in \cite{Riojas:2023pje}.} How does this interact with the AdS/CFT correspondence, and could the ``impact parameter" of a dual operator shine some light on the information paradox? In particular, what can be understood in terms of operators $\mathcal{O}$ with large (small) impact parameters? These operators would have either slowly (rapidly) oscillating sources with frequency $\omega$, or large (small) angular momentum $l$. Note these modes, if emitted near the horizon, would be trapped within the ``zone" of the black hole. This will surely be interesting to understand through future work.

\textit{Note Added}: During the preparation of this manuscript, \cite{Hashimoto:2023buz} appeared, which significantly overlaps with our results in some key areas. Their previous work on photon sphere images \cite{Hashimoto:2018okj,Hashimoto:2019jmw} was a primary source for this article and played a central role in our thought process. We had been interested in applying such observations to the Hawking--Page transition \cite{Hawking:1982dh,Witten:1998zw}, which the first-named author here has explored before in \cite{Karch:2023ekf}. We were guided by the notion that the transition occurs for global AdS black holes, which feature a photon sphere, and not for planar black holes, which do not. Conceptually, it seems that authors of \cite{Hashimoto:2023buz} are mainly interested in understanding QNMs and the emergent $SL(2,\mathbb{R})$ symmetries along the line of \cite{Hadar:2022xag}. Working in parallel, we were interested in exploring a possible connection between the AdS/CFT correspondence, the photon sphere, and the impact parameter defined for operators living on the boundary CFT. Both papers indicate the possibility of a large angular momentum phase space in AdS/CFT and point out that the analysis for QNMs in \cite{Cardoso:2008bp} applies under certain conditions, before performing a similar analysis in AdS spacetime. Note their approximate modes are more accurate for large overtone $n$. This is because they used the Posch-Teller potential, which smooths out the bottom of the hill in the effective potential at the photon sphere. 

\acknowledgments
We would like to thank Andreas Karch for many helpful and patient discussions, Aaron Zimmerman for clarifying some salient points about QNMs, and Matthew Dodelson and Cristoforo Iossa for bringing important and related earlier work \cite{Festuccia:2008zx} by Guido Festuccia and Hong Liu to our attention. We also thank David Berenstein, Giulio Bonelli, Minyong Guo, and Hongbao Zhang for helpful comments. The work of M.R. is supported, in part, by the U.S. Department of Energy under Grant-No. DE-SC0022021, an OGS Summer Fellowship, Dissertation Fellowship,  Dean's Strategic Fellowship, and a grant from the Simons Foundation (Grant
651440, AK), which supports the work of H.-Y.S. in part as well. H.-Y.S. also thanks the hospitality of Aspen Center for Physics, which is supported by National Science Foundation grant PHY-2210452.

\newpage
\appendix
\section{Placing 2nd Order ODEs in Schr\"odinger Form}
\label{sec:AppendixA}
This section reviews a standard method for converting 2nd order ODEs into standard form. Consider a one-variable 2nd order ODE:
\begin{equation}
    \phi''(x) + A(x) \phi'(x) + B(x) \phi(x) = 0.
\end{equation}
To obtain alternate forms of this expression, we simply define a new field:
\begin{equation}
    \psi(x) = e^{- \int P(x) dx} \phi(x).
\end{equation}
Then our differential equation becomes: 
\begin{equation}
    \psi''(x) + (A - 2P) \psi'(x) + (B - A P + P^2 - \partial_x P(x)) \psi(x) = 0
\end{equation}
While one can obtain many equivalent versions of the 2nd order ODE using this method, one can place it in the Schr\"odinger form by choosing $A = 2P$. This yields: 
\begin{equation}
    \psi''(x) + \left( B - \frac{1}{2} \partial_x A - \frac{1}{4} A^2 \right) \psi(x) = 0
    \label{eq:Schrodingerform2}
\end{equation}
Note that the derivative $\partial_x$ is taken with respect to the coordinate of $\phi(x)$. 

\subsection{Field Redefinitions and the Tortoise Coordinate}
In the case of the tortoise coordinates in Equation \eqref{eq:tortoise}, the derivative $\partial_x$ in Equation  is taken with respect to the original coordinate $r$. This changes the ODE from 
$$
    \left( \frac{d}{dr_*} \right)^2 \phi_l(r_*) + \left( \frac{2 F(r)}{r}\right) \left(\frac{d}{dr_*}\right)\phi_l(r_*) + \left(\omega^2 - \frac{l(l+1)F(r)}{r^2}\right)\phi_l(r_*) = 0,
$$
to a Schr\"odinger form given by:
\begin{equation}
    \left( \frac{d}{dr_*} \right)^2\psi_l(r_*) + \left( \omega^2 - \frac{l(l+1)F(r)}{r^2} - \frac{F(r)F'(r)}{r}\right)\psi_l(r_*) = 0,
    \label{eq:Schrodingerform3}
\end{equation}
where the field redefinition is given by: 
\begin{equation}
    \psi_l(r_*) = e^{\frac{1}{2} \int \frac{2 F(r)}{r} dr_*} \phi_l = e^{\int \frac{1}{r} dr} = r \phi_l(r_*),
    \label{eq:psiandphi}
\end{equation}
where $r$ is the \emph{regular} coordinate ``$r$", rather than the tortoise coordinate ``$r_*$''. 

\section{Review of the WKB Approximation}
\label{sec:WKB}
Here we briefly review the WKB approximation, cf. \cite{griffiths2018introduction}. The Schr\"odinger equation for a massive particle (setting $2m=\hbar = 1$ for simplicity) moving in a one-dimensional slowly-varying potential $V(x)$, with energy $\omega$, can be written as: 
\begin{equation}
    \left( \frac{d}{dx}\right)^2 \psi(x) = - p(x)^2 \psi(x),
\end{equation}
where the ``classical momentum'' is
\begin{equation}
    p(x) = \sqrt{\omega^2 - V(x)}.
    \label{eq:classical_momentum}
\end{equation}
The WKB method relies on the assumption that the amplitude, $A(x)$, and the rapidly oscillating phase, $\phi(x)$, are real and related to the  wavefunction as: 
\begin{equation}
    \psi(x) = A(x) e^{i \phi(x)}.
\end{equation}
Plugging this expression into the Schr\"odinger equation yields two expressions that simplify when the amplitude varies slowly. Standard arguments then yield: 
\begin{equation}
    \phi(x) = \int p(x) dx,
\end{equation}
\begin{equation}
    \psi(x) \approx \frac{1}{\sqrt{\abs{p(x)}}}e^{i \int p(x) dx}.
\end{equation}
For a sufficiently wide and essentially vertical barrier, the relative change in the amplitude is essentially given by the exponential of the integral of $p(x)$ over the barrier between the classical turning points for the barrier or potential well:
\begin{equation}
    \frac{\abs{\psi_{\text{in}}}}{\abs{\psi_{\text{out}}}} \sim e^{- \int \abs{p(x)}dx}.
\end{equation}
This approximation can be improved using the so-called connection formulas near the turning points:
\begin{equation}
\psi(x) \cong \begin{cases}\frac{2 D}{\sqrt{p(x)}} \sin \left[\frac{1}{\hbar} \int_x^{x_t} p\left(x^{\prime}\right) d x^{\prime}+\frac{\pi}{4}\right], & \text { if } x<x_t \\ \frac{D}{\sqrt{|p(x)|}} e^{-\frac{1}{\hbar} \int_{x_t}^x\left|p\left(x^{\prime}\right)\right| d x^{\prime}}, & \text { if } x>x_t\end{cases}
\end{equation}
where $x_t$ is the classical turning point in the potential, and $D$ is the normalization constant. When there is a vertical wall at $x=0$, one obtains the following:
\begin{equation}
\int_0^{x_t} p(x) d x=\left(n-\frac{1}{4}\right).
\end{equation}

Leaving out the conventional scenario discussed above, where frequency $\omega$ is treated as a real quantity in textbooks, it is natural to question the possibility of considering complex frequencies. This approach is not a novel concept, as it has been extensively employed when examining the QNMs of black holes, first pioneered in \cite{Schutz:1985km,Iyer:1986np,Iyer:1986nq} and thoroughly reviewed in \cite{Kokkotas:1999bd}. Its usage in AdS spacetimes was suggested by \cite{Berenstein:2020vlp}.

\section{Visualizing the Amplification Formula}
\label{sec:figures}
\begin{figure}
    \centering
    \includegraphics[scale=.55]{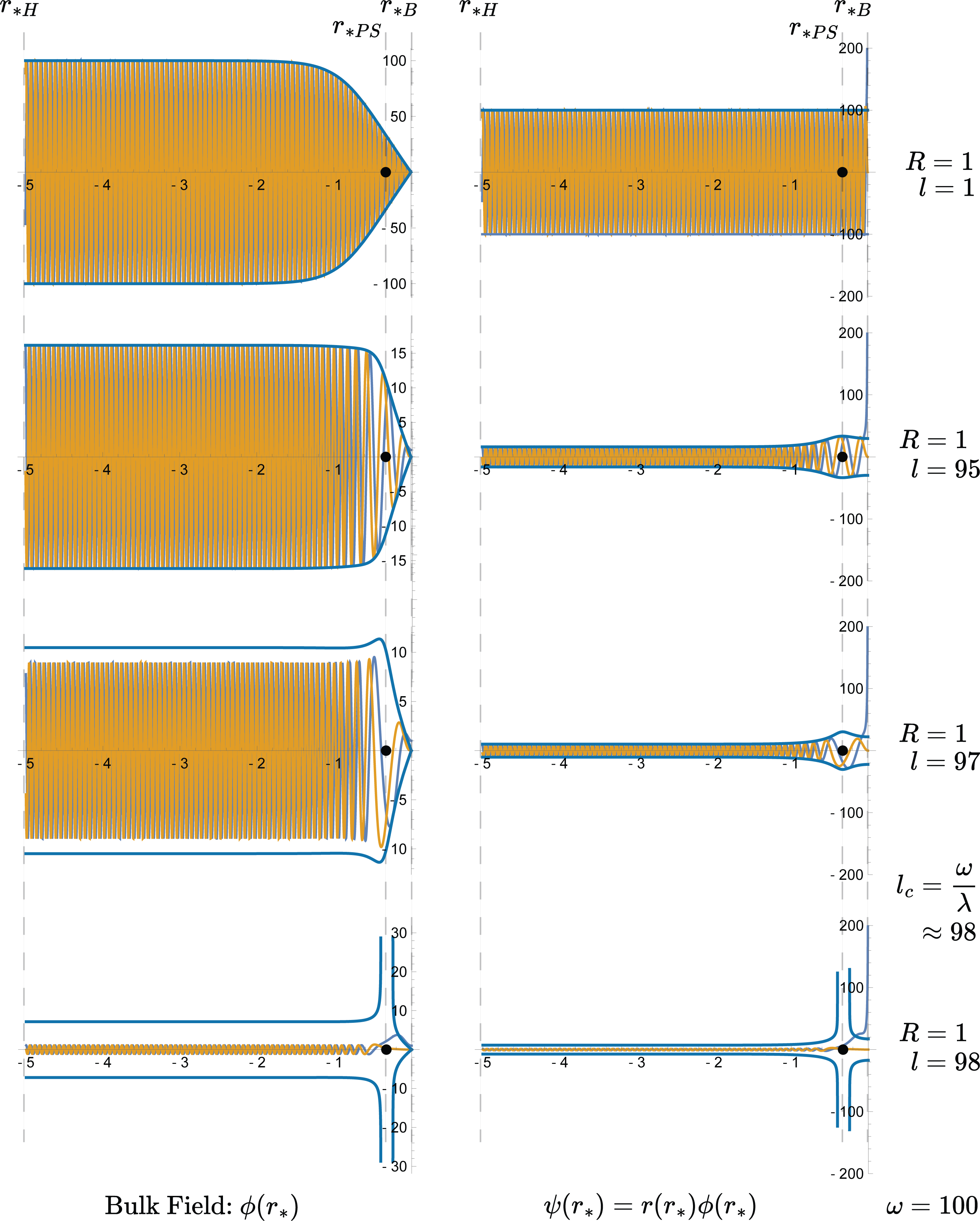}
    \caption{Here we present representative solutions to the differential equations for $\psi(r_*)$ and $\phi(r_*)$,
which satisfy the Schr\"odinger-type equation \eqref{eq:Schrodingerform1}, on the right, and the Helmholtz-type equation \eqref{eq:Helmholtzform}, on the left. The blue envelope is given by the amplification formula \eqref{eq:amplification_formula}. The black dot is the photon sphere location. This figure was generated at the Hawking-Page transition point $R=1$. For small $l$ the result resembles the top figure, where $\psi(r_*)$ is constant and division by the tortoise coordinate produces $\phi(r_*)$. The approximation breaks due to tunneling effects after the onset point $l_c = \frac{\omega}{\lambda}$. This transition, which is also where quasi-normal modes become available at $\omega = \lambda l$, is the same as the transition between massless particles which fall into the black hole and those which do not. Before the transition occurs, a bulge appears at the photon sphere because classical particles would orbit the photon sphere indefinitely at the onset point. If we take $R\rightarrow \infty$, the photon sphere approaches the AdS boundary and $\lambda \rightarrow 1$, so the transition disappears.}
    \label{fig:HP_amplitudes}
\end{figure}

\begin{figure}
    \centering
    \includegraphics[scale=.55]{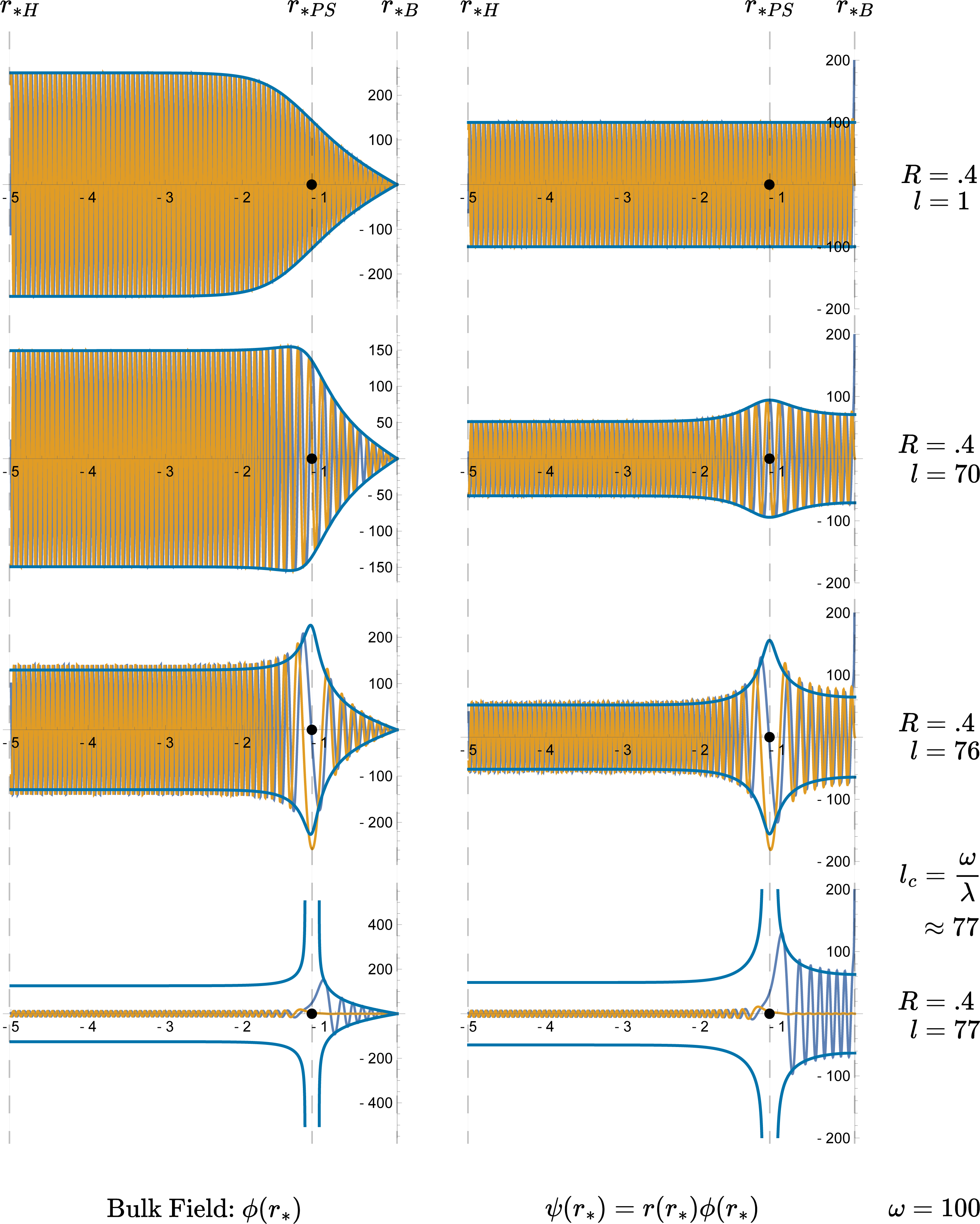}
    \caption{Here we present representative solutions to the differential equations for $\psi(r_*)$ and $\phi(r_*)$,
which satisfy the Schr\"odinger-type equation \eqref{eq:Schrodingerform1}, on the right, and the Helmholtz-type equation \eqref{eq:Helmholtzform}, on the left. The blue envelope is given by the amplification formula \eqref{eq:amplification_formula}. The black dot is the photon sphere location. This figure was generated for an unstable configuration,  below the Hawking-Page transition point, at $R=.4$. For small $l$ the result resembles the top figure, where $\psi(r_*)$ is constant and division by the tortoise coordinate produces $\phi(r_*)$. The approximation breaks due to tunneling effects after the onset point $l_c = \frac{\omega}{\lambda}$. This transition, which is also where quasi-normal modes become available at $\omega = \lambda l$, is the same as the transition between massless particles which fall into the black hole and those which do not. Before the transition occurs, a bulge appears at the photon sphere because classical particles would orbit the photon sphere indefinitely at the onset point.}
    \label{fig:small_amplitudes}
\end{figure}

\begin{figure}
    \centering
    \includegraphics[scale=0.55]{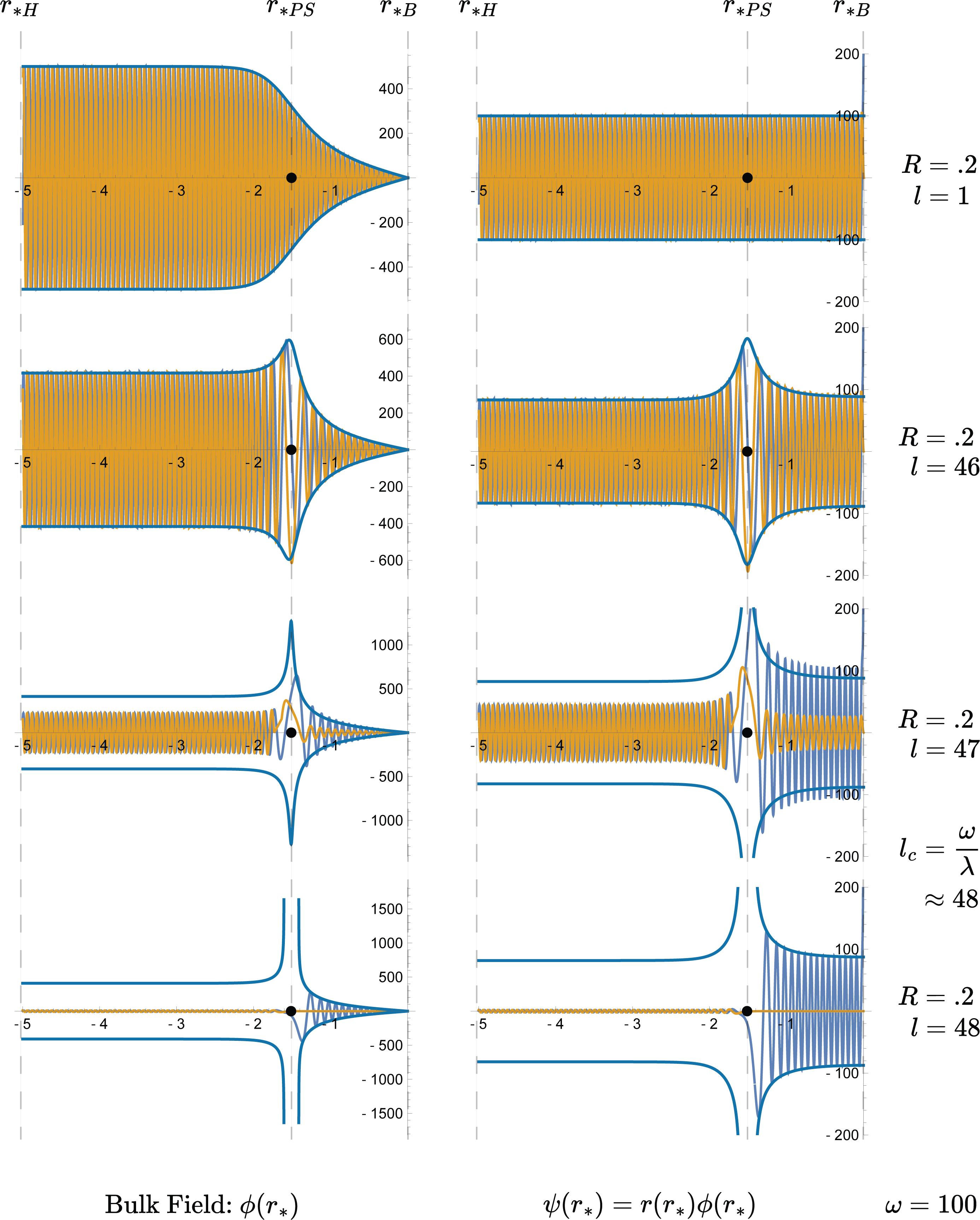}
    \caption{Here we present representative solutions to the differential equations for $\psi(r_*)$ and $\phi(r_*)$,
which satisfy the Schr\"odinger-type equation \eqref{eq:Schrodingerform1}, on the right, and the Helmholtz-type equation \eqref{eq:Helmholtzform}, on the left. The blue envelope is given by the amplification formula \eqref{eq:amplification_formula}. The black dot is the photon sphere location. This figure was generated for an unstable configuration, well below the Hawking-Page transition point, at $R=.2$. For small $l$ the result resembles the top figure, where $\psi(r_*)$ is constant and division by the tortoise coordinate produces $\phi(r_*)$. The approximation breaks due to tunneling effects after the onset point $l_c = \frac{\omega}{\lambda}$. This transition, which is also where quasi-normal modes become available at $\omega = \lambda l$, is the same as the transition between massless particles which fall into the black hole and those which do not. Before the transition occurs, a bulge appears at the photon sphere because classical particles would orbit the photon sphere indefinitely at the onset point.}
    \label{fig:tiny_amplitudes}
\end{figure}

Since Figures \ref{fig:HP_amplitudes}-\ref{fig:tiny_amplitudes} play a significant role in our presentation, and take up a considerable amount of space, we have collected them in this separate appendix for convenience. 

\newpage

\bibliographystyle{JHEP}
\bibliography{Main}
\end{document}